\documentclass[twoside,11pt]{article}
\usepackage{obs_study_style}

\heading{}{}{}{}{}{Parikh et al}
\usepackage{hyperref}
\usepackage{amsmath}
\usepackage{multirow}
\usepackage{color}
\usepackage{tikz}
\usepackage{graphicx}
\usepackage{booktabs}
\usepackage{ulem}
\usepackage{float}
\usepackage{lscape}
\usepackage{rotating}
\usepackage{longtable}
\usepackage{etoc}
\usepackage{appendix}

\newcommand{\red}[1]{{\color{red}\emph{xx #1 xx}}}

\newcommand{\gdr}[1][t]{g_{DR}(#1, \bX)}
\newcommand{\g}{\hat{g}(t, \bX)}
\newcommand{\Tit}{T_i(t)}
\newcommand{\Om}{\Omega}
\newcommand{\om}{\omega}
\newcommand{\ind}{\mathbb{I}}
\newcommand{\E}{\mathbb{E}}
\newcommand{\V}{\mathbb{V}}
\newcommand{\bx}{\mathbf{x}}
\newcommand{\bX}{\mathbf{X}}
\newcommand{\bY}{\mathbf{Y}}
\newcommand{\cX}{\mathcal{X}}
\newcommand{\cY}{\mathcal{Y}}
\newcommand{\cS}{\mathcal{S}}
\newcommand{\R}{\mathbb{R}}
\newcommand{\indep}{\mathrel{\text{\scalebox{1.07}{$\perp\mkern-10mu\perp$}}}}
\DeclareMathOperator*{\argmin}{arg\,min}
\DeclareMathOperator*{\argmax}{arg\,max}
\newcommand{\note}[1]{{\color{red}[\textbf{Note:} #1]}}
\newcommand{\mut}[2]{{\mu(#1, t, #2)}}
\newcommand{\muitobs}{\mut{\bx_i}{0}}
\newcommand{\muitexp}{\mut{\bx_i}{1}}
\newcommand{\mutobs}{\mut{\bx}{0}}
\newcommand{\mutexp}{\mut{\bx}{1}}

\newcommand{\et}[2]{{e(#1, t, #2)}}
\newcommand{\pq}[1]{\left\Vert#1\right\Vert_{P, q}}
\newcommand{\ptwo}[1]{\left\Vert#1\right\Vert_{P, 2}}
\newcommand{\eitobs}{\et{\bx_i}{0}}
\newcommand{\eitexp}{\et{\bx_i}{1}}
\newcommand{\etobs}{\et{\bx}{0}}
\newcommand{\etexp}{\et{\bx}{1}}
\newcommand{\pri}{p(\bx_i)}
\newcommand{\pr}{p(\bx)}

\newcommand{\muitobshat}{{\hat{\mu}(\bx_i, t, 0)}}
\newcommand{\muitexphat}{{\hat{\mu}(\bx_i, t, 1)}}
\newcommand{\eitobshat}{{\hat{e}(\bx_i, t, 0)}}
\newcommand{\eitexphat}{{\hat{e}(\bx_i, t, 1)}}
\newcommand{\prihat}{\hat{p}(\bx_i)}

\newcommand{\thetahat}{\hat{\theta}}
\newcommand{\Psiithat}{\widehat{\phi}_i(t)}
\newcommand{\Psiit}{\phi_i(t)}
\newcommand{\etab}{\boldsymbol{\eta}}
\newcommand{\etabhat}{\hat{\boldsymbol{\eta}}}
\newcommand{\sigmasq}{\sigma^2}
\newcommand{\sigmasqhat}{\hat{\sigma}^2}
\newcommand{\Hzerot}{\mathbb{H}^0(t)}
\newcommand{\eps}{\epsilon}
\newcommand{\epsy}{\mathcal{E}_Y}
\newcommand{\alphux}{\alpha(\bX)}
\newcommand{\gamux}{\gamma(\bX)}
\newcommand{\betux}{\beta(\bX)}
\newcommand{\alphuxz}{\alpha_0(\bX)}
\newcommand{\gamuxz}{\gamma_0(\bX)}
\newcommand{\betuxz}{\beta_0(\bX)}
\newcommand{\EOm}{\mathbb{E}_{\Omega}}
\newcommand{\POm}{\mathrm{P}_{\Omega}}
\newcommand{\tauobs}{{\tau^{obs}}}
\newcommand{\tauobshat}{{\hat{\tau}^{obs}}}
\newcommand{\nuhat}{\hat{\nu}}
\newcommand{\Lambdaithat}{{\widehat{\Lambda}_i(t)}}
\newcommand{\Lambdait}{{\Lambda_i(t)}}
\newcommand{\phat}{{\hat{p}}}
\newcommand{\gammasq}{\gamma^2}
\newcommand{\gammasqhat}{\hat{\gamma}^2}

\ShortHeadings{Integrating Experimental and Observational Studies}{Parikh, Morucci, Orlandi, Roy, Rudin and Volfovsky}
\firstpageno{1}

\begin{document}

\title{A Double Machine Learning Approach for Combining Experimental and Observational Studies}

\author{
    \name Harsh Parikh\footnote{Co-first Authors} \email harsh.parikh@yale.edu \\
       \addr Department of Biostatistics\\
       Yale University\\
       United States of America
       \AND
       \name Marco Morucci$^*$
       \email moruccim@msu.edu \\
       \addr Department of Political Science\\
       Michigan State University \\
       United States of America
       \AND
       \name Vittorio Orlandi$^*$ \email vittorio.orlandi@duke.edu \\
       \addr Department of Statistical Science\\
       Duke University \\
       United States of America
       \AND
       \name Sudeepa Roy \email sudeepa.roy@duke.edu \\
       \addr Department of Computer Science\\
       Duke University \\
       United States of America
       \AND
       \name Cynthia Rudin \email cynthia.rudin@duke.edu \\
       \addr Department of Computer Science\\
       Duke University \\
       United States of America
       \AND
       \name Alexander Volfovsky \email alexander.volfovsky@duke.edu \\
       \addr Department of Statistical Science\\
       Duke University \\
       United States of America}

\maketitle

\begin{abstract}%
Experimental and observational studies often lack validity due to untestable assumptions. We propose a double machine learning approach to combine experimental and observational studies, allowing practitioners to test for assumption violations and estimate treatment effects consistently. Our framework proposes a falsification test for external validity and ignorability under milder assumptions. We provide consistent treatment effect estimators even when one of the assumptions is violated. However, our no-free-lunch theorem highlights the necessity of accurately identifying the violated assumption for consistent treatment effect estimation. Through comparative analyses, we show our framework's superiority over existing data fusion methods. The practical utility of our approach is further exemplified by three real-world case studies, underscoring its potential for widespread application in empirical research.
\end{abstract}

\begin{keywords}
    Data Fusion, Generalizability, External Validity, Observational Study, Machine Learning 
\end{keywords}

\section{Introduction}
Experiments and observational studies are both indispensable for estimating treatment effects and investigating causal questions of interest. Experiments, or randomized control trials (RCTs), guarantee unbiased estimation of sample average treatment effects via the randomization of treatment across the experimental population. However, they are typically small, expensive, and slow to design and implement. Another concern involves the experimental units themselves, which may be recruited via a non-randomized process and therefore may not be representative of the superpopulation of interest; in such cases, we would say that the experiment lacks external validity. If this is the case, the conclusions of the experiment may not be generalizable to the superpopulation, which is oftentimes of primary interest. On the other hand, observational studies, in which units self-select into treatment, tend to be larger and include units that are representative of the superpopulation. However, analysis of observational studies is plagued by observed and unobserved confounders -- covariates that affect both units' propensity to self-select into treatment and their response to that treatment. While a variety of methods -- matching and weighting being two large classes among them -- can be used to adjust for confounders, they typically rely on the crucial assumptions that they can accurately model the treatment-covariate or response-covariate relationships and that the available covariates are sufficient for doing so. This latter assumption, which is often referred to as strong or \textit{conditional ignorability}, is untestable using observational data alone \citep{rosenbaum1983assessing}.

In this paper, we propose methods for jointly using experimental and observational data to (i) test for violations of at least one of validity and ignorability and (ii) estimate population average treatment effects even under such violations. To the best of our knowledge, no formal testing framework has been proposed for these violations, and as we discuss below, existing estimation approaches target similar estimands by assuming external validity and conditional ignorability. Crucially, our estimators are valid under violations of certain sets of assumptions that are often required in the literature, making our methods applicable in settings where other methods may be inappropriate. We provide a detailed discussion of the literature in Section~\ref{sec:lit}. We derive the influence functions for our scenario of interest. Further, we rigorously show that our estimators possess rate-double-robustness properties, ensuring their root-n consistency, asymptotic normality, and unbiasedness in Sections~\ref{sec:testing}~and~\ref{sec:drate}. The properties of these estimators allow us to address testing as well as estimation tasks. 
We emphasize that we accomplish these tasks by making use of both observational \textit{and} experimental data, despite the fact that one of external validity or conditional ignorability may be \textit{violated}. 

After establishing the theoretical properties of our tests and estimators, we apply them to three real-world studies -- one of them is in Section~\ref{sec:star} while the other two are relegated to supplementary material. In doing so, we showcase the utility of our method for practitioners using observational and experimental data to answer their causal questions of interest. The analyses reach different conclusions about violations of ignorability and validity, illustrating the wide range of data that can be appropriately analyzed with our methods. In the first, we reanalyze data from Project STAR  \citep[Student-Teacher Achievement Ratio;][]{project_star_data,mosteller2014tennessee}, which consists of experimental and observational data gathered to study the impact of small class size in early schooling on future test scores. We estimate a small, positive treatment effect (about 6 points on a test that ranges between 400-800 total points) attributable to a small class size on a standardized test, consistent with prior results on this dataset. 
Leveraging the theory developed in our paper, we can point-identify the amount of selection bias in the observational data using breakdown-frontiers analysis \citep{masten2020inference}. Our estimates indicate that an unobserved confounder may have led to a selection bias of 16 points in the observational study. Second, we analyze data gathered by the Coronary Artery Surgery Study (CASS) \citep{CASS}, sponsored by the National Heart, Lung, and Blood Institute, to compare the effects of coronary artery bypass surgery to non-surgical medical treatment. Here, our population ATE estimate is consistent with those in past literature, as is the fact that we find no evidence of unobserved confounding or violations of external validity. Lastly, we explore the application of our approach to the experimental data from the National Supported Work Demonstration (NSW) and observational controls from the Population Survey of Income Dynamics (PSID) \citep{lalonde1986,dehejia2002}. Our analysis indicates that external validity is violated in the NSW sample. Interestingly, we find that an existing approach to treatment effect estimation proposed by \cite{malts} can prune heavily confounded units and recover the experimental ATE using observational data.

The rest of the paper proceeds as follows. In Section \ref{sec:lit}, we discuss related work. In Section \ref{sec:prelim}, we introduce our setting, notation, and various quantities of interest. Sections \ref{sec:testing} and \ref{sec:drate} outline our procedures for testing and estimation. Sections \ref{sec:star} \& \ref{sec:synth_exp} apply our methods to the Synthetic and STAR datasets. Further, Appendices~\ref{sec:cass} \& \ref{sec:lalonde} discusses the application of our approach to CASS and Lalonde datasets. Section \ref{sec:conclusion} summarizes our contributions and discusses future extensions.
\section{Literature Review}\label{sec:lit}

The advent of huge observational datasets has led to the rapid growth of several (mostly disjoint) subfields.
In this section we aim to bridge those together and place our proposed methodology within this broader context. We concentrate on the following fields: (i) generalizability, (ii) combining observational and experimental datasets, (iii) violations of ignorability, (iv) negative controls, and (v) double robustness.

\textbf{Generalizability.} Though mild conditions ensure that treatment effects estimated from randomized control trials are unbiased, this unbiasedness is with respect to the distributions within the experiment; that is, the unbiasedness only necessarily holds for the average treatment effect \textit{within} the experimental sample. One branch of work on generalizability defines the conditions under which effects can be appropriately generalized to larger populations. For example, \cite{egami_hartman_ex_validity} categorize the types of validity that are necessary for an experiment's results to be appropriately extended to the population.
Furthermore, they specify assumptions leading to these forms of validity and show identification of population treatment effects given they hold. Other work on generalizability deals with adjusting treatment effects derived from an experimental sample to extend them to the population. 
\citet{sens_w_bias_funs}, for example, posit bias functions describing how lack of exchangeability between the units in experiment and population affects estimates, and they use these bias functions to perform sensitivity analysis. If the bias function is correctly specified, standard weighting, outcome-based, and doubly robust estimators can be modified to correctly debias estimates. \citet{pop_outcomes_adjust_exp} use population data to learn outcome models that are used to estimate and residualize experimental outcomes; the residuals are used in conjunction with sampling weights to estimate the population average treatment effect (PATE). 

\textbf{Combining Experimental and Observational Studies.}  
Observational and experimental data have recently been used together to improve the efficiency of treatment effect estimation procedures. In this setting, the idea is to take advantage of both the internal validity of the experimental sample and the external validity (and typically larger sample size) of the observational data. This field has grown rapidly in the past several years and here we briefly mention some key work. {For a broad review of this literature, see \cite{brantner2023methods} and \cite{colnet2020causal}.}

\cite{design_exp_from_obs} use observational data to estimate confidence sets for the variance of within-strata potential outcomes and incorporate the sets into a regret minimization problem used to design an experiment. Conversely, \cite{emulate_rct} seek to duplicate randomized control trials, isolating trial-mimicking populations and then conducting propensity score matching therein. \cite{shrinkage_combine} construct a James-Stein type shrinkage estimator that, under mild conditions, is guaranteed to outperform estimates constructed using only one of the two data sources. Similarly, \cite{pscore_combine} use a propensity score model trained on the observational data to stratify experimental data based off what their propensity scores would have been had they been in the other sample. The resulting stratified design is then used to estimate treatment effects using both samples. \cite{hartman_sekhon} discuss the assumptions required for identification of the population average treatment effect on the treated (ATT) from experimental data and suggest a method for using baseline characteristics and outcomes available from non-randomized data to adjust experimental estimates appropriately. \cite{long_term_combine} focus on the related problem of estimating treatment effects on a long-term outcome that is only observed in the experimental data. \cite{ts_combine} use time-series data to learn a linear outcome model for each individual observational unit; the estimated parameters can then be incorporated into a model for the experimental data that may increase power of the ensuing analysis. \cite{mutual_debias} propose a multi-study instrumental variable approach. They show how the approach allows for nonparametrically debiasing both experimental (biased due to lack of external validity) and observational estimates (biased due to unobserved variables affecting selection into treatment); in practice, an instrumented difference-in-differences is used to point-identify local average treatment effects. \cite{Triantafillou2021} use experimental data to learn feature sets that should be adjusted for when estimating treatment effects on observational data. \cite{Ghassami2022} focus on short- and long-term outcomes and proposes various assumption sets under which causal effects on the latter can be identified. \cite{Dang2022} estimate treatment effects using negative control outcomes -- outcomes that are unaffected by treatment, but affected by some unobserved confounder induced bias. \cite{power_likelihood} augment experimental data with observational data, where the likelihood of the observational data is raised to some power $\eta < 1$ and $\eta$ is chosen to maximize the expected log pointwise predictive density on the experimental data. Likelihood-based inference is then performed to estimate treatment effects.

Our work broadens many of the above works in a fundamental way: while we also take advantage of having both observational and experimental data to better estimate population average treatment effects, we can do so in the presence of unobserved confounding, and even formally test for its impact on the treatment selection mechanism in the observational sample. Furthermore, our estimators are computationally efficient and do not depend on the presence of specialized structure in the data (e.g., existence of instrumental variables or proxy outcomes).

\textbf{Violation of Ignorability.} Our work is also related to the study of the impact of unobserved confounders on a treatment effect estimation procedure. In settings in which only observational data are available, most works (explicitly or implicitly) make an assumption equivalent to strong ignorability to identify causal quantities of interest. This is because, in many of these scenarios, it is impossible to identify or estimate the degree of unobserved confounding using observational data alone. One approach to such settings relies on sensitivity analyses to potential unobserved confounding \citep{rosenbaum_book, intro_to_sa}. Such analyses typically posit parametric models (but see, e.g., \cite{sa_wo_assumptions} for a nonparametric approach) for how an unobserved confounder affects treatment and outcome and then assess how large the effect of the unobserved confounder needs to be to change the treatment effect estimate by some specified amount. Crucially, however, such methods cannot suggest anything about the true degree of unobserved confounding present. In general, even if a sensitivity model is correctly specified, extreme sensitivity to the specified amount of unobserved confounding does \textit{not} imply a high degree of unobserved confounding; neither does a lack of sensitivity imply a low degree. 

In experimental settings, the randomization of treatment protects against any unaccounted selection-bias. There is currently a dearth of work concerning the study of unobserved confounders using both experimental and observational data. \citet{kallus_unobserved} consider the problem of CATE estimation in this setting. Theoretical results rely upon assumptions of external validity and of a parametric structure for the impact of the unobserved confounder. Furthermore, the validity of the method depends on an identification assumption whose form depends on the posited parametric form. Lastly, it is not clear how this approach generalizes to estimation of average treatment effects. \citet{shrinkage_combine} suggest a sensitivity analysis that yields an `implied' degree of unobserved confounding, based off an augmented inverse propensity weighted estimator. The method is restricted to this estimator, however, and, as with other sensitivity analyses, cannot truly identify the presence of an unobserved confounder. \citet{iv_test_unobserved} describes a set of sufficient conditions on an instrument that allows for testing of unobserved confounding, though their method does not allow for identification of average causal effects under the presence of confounders. 

We further draw connections to the literature on negative control and double robustness.

\textbf{Negative Controls.} Negative controls are variables that do not causally affect the outcome of interest and thus have a null effect by design \citep[see][for detailed review]{lipsitch2010negative,tchetgen2020introduction}, and have recently been used in complex causal inference settings. For example, unobserved confounders are integral to the notion of negative control outcomes in the work of \citet{Dang2022}. However, the required dependencies between the unobserved confounder and the treatment, outcome, and negative control outcome must be assumed. Our work can also be interpreted in this context: Here, the variable informing about a unit's inclusion in an experimental sample can be considered as a negative control under the assumption that selection into an experimental sample has no direct causal path to the outcome. The key difference between negative control literature and our work here is that for units with $S=1$, the treatment assignment is independent of any unobserved confounders. The knowledge of the treatment assignment mechanism in the experiment not only allows testing for unobserved confounding in the corresponding observational study and external validity of the experimental sample but also estimation of population ATE even if one of these assumptions is violated.

\textbf{Double Robustness.} Our work also bears a resemblance to the literature on the doubly-robust estimation of causal effects \citep{robins_original_dr, summary_dr}, where outcome and treatment-selection estimators are combined to create an estimator consistent under misspecification of either propensity or prognostic score models (but not both simultaneously). Our work distinguishes itself from the above in we consider violations of assumptions, not inappropriate model specifications. We will later show that our estimator is also doubly robust in the traditional sense; however, this is not the primary focus of this paper.

\section{Preliminaries}\label{sec:prelim}
\begin{figure}
    \centering
    \includegraphics[width=0.7\textwidth]{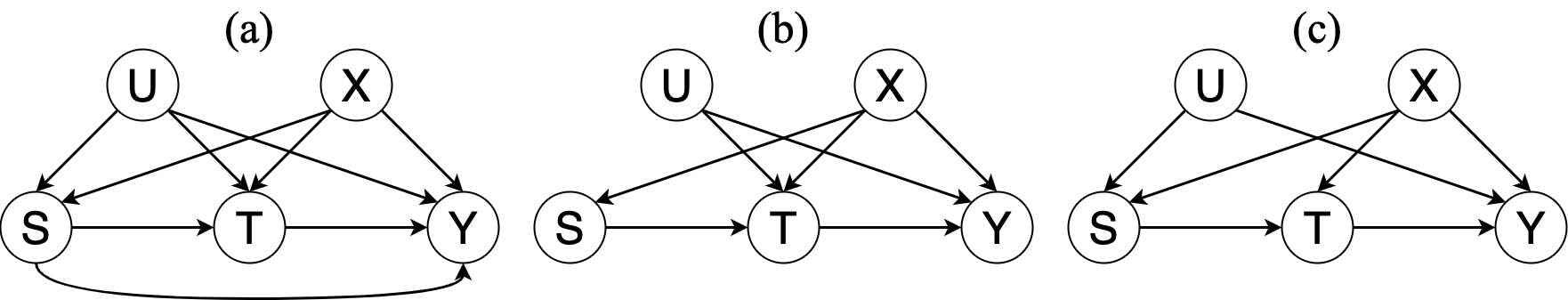}
    \caption{Causal DAGs. Panel (a): potential relationships without any assumptions. Panel (b): variable relationships under A1, A2, A3, and A5 while A4 is explicitly violated. In this case, TE identification in the experimental sample can be understood as a special case of instrumental variable identification where $S$ is the instrument. Panel (c): variable relationships under A1, A2, A4, A5, while A3 is explicitly violated. In this case, the relationship between U, T, and Y is the same in both the experimental and observational populations.}
    \label{fig:expob_causalDAG}
\end{figure}
\subsection{Causal Inference: Notation}\label{subsec:notation}
We consider settings with $n$ observed units, indexed by $i$, where we are interested in estimating the causal effect of a binary treatment $T$ on outcome $Y$. Each unit is associated with two real-valued potential outcomes, $Y_i(0)$ and $Y_i(1)$ corresponding to two treatment choices. While we assume binary treatment for simplicity, our results immediately extend to the cases of $M > 2$ potential outcomes. We take $Y_i(t) \in \cY$ for $t \in \{0, 1\}$, where $\cY \subseteq \R$ that we will further characterize shortly. Each unit is also assigned a treatment whose observed value is denoted by $T_i \in \{0, 1\}$. The variable $Y_i$ (note the absence of parentheses) denotes the observed outcome of unit $i$, which we assume to be the only potential outcome corresponding to their assigned treatment, i.e., $Y_i = Y_i(T_i)$. In addition, we also observe, for each unit, a vector of $d$ pre-treatment covariates $\bX_i \in \cX$, such that $\cX$ is a compact subset of $\R^d$. 

In our setting, the $n$ units are divided into two subsets: a set of \textit{observational} units and a set of \textit{experimental} units. We will use the variable $S_i = 1$ to denote unit $i$ is in the experimental sample and $S_i = 0$ to denote unit $i$ is in the observational sample. As we will shortly see, the statistical behavior of the variables associated with a unit in the observational set will be different than that of a unit in the experimental set. 
Lastly, we consider an unobserved confounder $U \in \mathcal{U} \subseteq \R$ that may causally affect $T$, $S$, and $Y$.

Throughout the rest of this paper, we will use capitalized letters to represent random variables, lowercase letters for values within the domain of the distribution of these random variables, and $f_A$ and $F_A$ to indicate the probability density function (pdf) and cumulative distribution function (CDF) of the random variable $A$, respectively. We will be studying several functionals of the joint distribution of  $Y$, $X$, $T$, and $S$. We introduce some shorthand notation for these quantities of interest here:
\begin{align}
    \mutobs &:= \E[Y|\bX = \bx, T = t, S = 0] \label{eq:muobs}\\
    \mutexp &:= \E[Y|\bX = \bx, T = t, S = 1]\label{eq:muexp}\\
    \etobs &:= \Pr(T = t|\bX = \bx, S = 0)\label{eq:eobs}\\
    \etexp &:= \Pr(T = t|\bX = \bx, S = 1)\label{eq:eexp}\\
    \pr &:= \Pr(S = 1|\bX = \bx)\label{eq:pri}.
\end{align}
with corresponding estimates denoted by hats: $\hat{\mu}(\bx, t, 0), \hat{e}(\bx, t, 1), \hat{p}(\bx)$, etc.

The first two quantities are the conditional expectations of the outcome given covariates $\bX$  and treatment $t$ in the observational (Equation~\eqref{eq:muobs}) and experimental samples (Equation~\eqref{eq:muexp}), respectively. The following two quantities are the propensity scores for treatment level $t$ also in the observational (Equation~\eqref{eq:eobs}) and experimental samples (Equation~\eqref{eq:eexp}), respectively. Finally, $\pri$ denotes the probability that unit $i$ is assigned to the experimental sample, conditionally on its observed covariates (Equation~\eqref{eq:pri}). All these quantities are unknown, but estimable with the observed data. In an ideal experimental setting, both $\eitexp$ and $\pri$ would be known by the analyst, but this need not be the case in our framework. We now discuss assumptions key to our framework and that will help us estimate the above quantities.

\subsection{Causal Inference: Assumptions}
Here, we state several assumptions that are common in the causal inference literature and which are of key importance in our work. In our setting, several of these become testable and certain violations do not hinder consistent estimation of causal effects.

\noindent\textbf{A1 (Data Distribution)} For notation's sake, define the random vector $\Om := (Y, T, S, \bX)$. Let $f_{\Om}$ be a probability distribution over the joint domain $\cS = \cY \times \{0, 1\} \times \{0, 1\} \times \cX$. We assume that the full data are iid random samples from this distribution, i.e.: $\{\Om_i\}_{i=1}^n \overset{iid}{\sim} f_{\Om}$, and that $\Om$ (note the absence of indices) denotes an arbitrary draw from $f_{\Om}$. We also maintain the following standard overlap assumption: for all $\om \in \cS$, we assume $0 < f_{\Om}(\om) < 1$. \\
\textbf{A2 (Internal Validity of the Experiment)} We assume that $(Y(0), Y(1)) \indep T \mid \bX=\bx, S = 1$, i.e., if $i$ is an experimental unit, then its treatment assignment is independent of its potential outcomes conditional on its covariates.  
\\\textbf{A3 (External Validity of the Experiment)} Adjusted for pre-sampling covariates $\bX$, the conditional means of potential outcomes are exchangeable across study populations: $\E[Y(t) \mid \bX=\bx, S=1] = \E[Y(t) \mid \bX=\bx, S=0]$ for all $t$ and $\bx$.\\
\textbf{A4 (Conditional Ignorability)} Adjusted covariates $\bX$, in the observational study, the conditional means of potential outcomes are exchangeable across treatment arms: 
$\E[Y(t) \mid \bX=\bx, T=1, S=0] = \E[Y(t) \mid \bX=\bx, T=0, S=0]$ for all $t$ and $\bx$.
\\\textbf{A5 (Sampling Ignorability)} For all $t$ and almost surely over $f_{\Omega}$ we have: $Y \indep S \mid \bX = \bx, T=t, U=u$, i.e., being sampled into the experimental set is independent of the outcome conditionally on the covariates, unobserved confounder, and the treatment level.  
\\\textbf{A6 (Nuisance Parameter Estimation)}
For all $t \in \{0, 1\}$ and $s \in \{0, 1\}$:
\begin{gather}
     \ptwo{\hat{\mu}(\bx, t, s) - \mu(\bx, t, s)} \times \ptwo{\hat{e}(\bx, t, s) - e(\bx, t, s)} = o(n^{-1/2}) \label{eq:nuis_est1}\\
     \ptwo{\hat{\mu}(\bx, t, s) - \mu(\bx, t, s)} \times \ptwo{\hat{p}(\bx) - p(\bx)} = o(n^{-1/2})\label{eq:nuis_est2}
\end{gather}
where $\pq{f} = \pq{f(\Omega)} = (\int |f(\omega)|^qd {P}(\omega))^{1/q}$ denotes the $L^q(P)$ norm with respect to the distribution of the data. 

%

\paragraph{\textbf{Discussion of Assumptions}} Assumption A1 is the standard overlap assumption and a regularity condition required to operationalize our double machine learning procedure. 
Assumption A2, which states that experimental units are assigned treatment effectively at random conditional on the observed covariates, should hold in most cases where an experiment is properly conducted and treatment truly is randomized. As a consequence of A2, we have $P(Y(t) | \bX =\bx, T=t, S=1) = P(Y(t) | \bX =\bx, S=1)$. It may still be violated, however, in certain cases, such as if treated units with higher potential outcomes tend to drop out of the experiment and are excluded from further analysis. The knowledge of a unit's treatment status would give information about their potential outcomes and vice versa, violating A2. 

Assumption A3 states that units are selected into the experimental set effectively at random conditional on the observed covariates. Importantly, if there exists an unobserved confounder that is not marginally independent of the potential outcomes, it must be marginally independent of $S$. A practical example of a setting where A2 and A3 both hold (barring violations of A2 akin to the above) is one in which a researcher first samples $n$ individuals from a superpopulation of interest, measures baseline covariates, $\bX$, and chooses a subset to invite into a lab experiment based on $\bX$, while the rest of the individuals are allowed to self-select into treatment. This is approximately the setup of the CASS study we will analyze later (Section \ref{sec:cass}). As a consequence of A3, we have $P(Y(t) | \bX =\bx, S=s) = P(Y(t) | \bX =\bx)$.

Assumption A4 is the standard conditional ignorability assumption and asserts that there are no unobserved confounders affecting treatment and outcome. This assertion is untestable from observational data alone. However, A4 is typically assumed by default in observational analyses (though sensitivity analyses may be performed to assess the sensitivity of causal estimates to varying strengths of a posited unobserved confounder). As a consequence of A4, we have $P(Y(t) | \bX =\bx, T=t, S=0) = P(Y(t) | \bX =\bx, S=0)$.
Later, we will show that having experimental data makes this assumption testable. 

Assumption A5 assumes that sampling into the experiment does not have a direct causal path affecting the outcome i.e., in Figure~\ref{fig:expob_causalDAG}(a) there is no arrow from S to Y. Such an assumption would be unreasonable if participation in the experiment \textit{in and of itself} changed a unit's outcome. For example, consider an experiment in which individuals are paired and their conversations monitored for indications of politeness. If knowledge of being monitored leads one to speak more politely than they would otherwise, A5 would be violated. This assumption can also be seen as a form of the consistency or SUTVA assumption extended to our two population settings. 

Assumptions A3 to A5 (or a version of them) are common across approaches that combine experimental and observational studies to estimate treatment effects \citep{rudolph2023improving, lu2019causal, wu2022integrative}. Further, \cite{wu2022integrative} considers estimating effects in a scenario where A4 is violated.

Lastly, Assumption A6 specifies the rates at which we must be able to estimate nuisance parameters. Importantly, achieving these rates does \textit{not} require correct specification of a parametric model and can be accomplished with flexible nonparametric or machine learning methods. These rate conditions arise naturally from efficient influence function based estimators in semiparametric statistics, which underpins the construction of doubly robust and orthogonal estimators \citep{bickel1993efficient,robins1994estimation,rotnitzky2012improved,farrell2015robust,chernozhukov2018double,kennedy2024semiparametric}.


Importantly, the above assumptions imply that some of our quantities of interest have the following key properties:
\begin{align}
    \mutexp &= \E[Y(t)|\bX = \bx,S=1] &  (\text{by A2}) \label{eq:6}\\
    \etexp &=  \Pr(T=t|\bX=\bx, Y(t)=y(t), S = 1)  &  (\text{by A2}) \label{eq:7}\\
    \pr &=  \Pr(S=1|\bX=\bx, Y(t)=y(t)).  &  (\text{by A3}) \label{eq:8}
\end{align}
Equation~\eqref{eq:6} is implied by A2, and is key in any causal inference framework as it permits identification of the conditional mean: potential outcomes can be consistently estimated for every experimental unit. Equation~\eqref{eq:7} is also directly implied by A2 and is a consequence of the fact that assignment to treatment in the experiment does not depend on a unit's potential outcomes. Finally, by A3, Equation~\eqref{eq:8} states that assignment to the experimental sample is independent of the potential outcomes, conditionally on covariates. Note that all these quantities are guaranteed to be non-degenerate by the distributional requirements in A1.

\section{Impossibility of Double Resilience}

We begin our exposition with a result that motivates the estimators we propose in the rest of our paper: while it is possible to leverage experimental and observational data to overcome violations of A3 or A4, \textit{combining experimental and observation data without the knowledge of violations of A3 and/or A4 may not yield the expected benefits and can result in a biased estimate}. 

This is important as it clarifies the requirements of estimation with experimental and observational data in a very general sense: users of these methodologies must possess knowledge of whether the experimental data generalize, or the observational set satisfies unconfoundedness, and knowledge of the fact that either of these assumptions may hold is not sufficient.  

In order to prove the fact that knowledge of whether A3 or A4 is violated is necessary for estimation, we investigate whether an estimator for treatment effects might exist that is efficient under violations of either A3 or A4. We would call such an \textbf{estimator} \emph{doubly resilient} as it would be resilient to violation of one of two assumptions. The terminology `double resiliency' contrasts with the term `double robustness'; the latter refers to the misspecification of models, not the violation of assumptions. We formally define this property below:

\begin{definition}[Double Resilience] Given A1, A2, and A5, a doubly resilient estimator $\gdr[1]-\gdr[0]$ is an unbiased and consistent estimator of conditional average treatment effects if at least one of A3 or A4 is \textbf{not} violated. 
More precisely, we say that $\gdr$ is doubly resilient if under assumptions (A1, A2, A3, A5) or under assumptions (A1, A2, A4, A5), then $\forall t,\bX,s$:  $\mathbb{E}[\gdr]  = \mathbb{E}\left[ Y(t) \mid \bX \right]$ and $\mathbb{E}[\gdr \mid S=s]  = \mathbb{E}\left[ Y(t) \mid \bX, S=s \right]$.
\end{definition}

Importantly, we have the following result:
\begin{theorem}[Doubly Resilient Estimators Do Not Exist]\label{thm:impossibility}
    There does not exist any doubly resilient estimator $\gdr$.
\end{theorem}
The proof can be found in the supplement. The intuition behind the proof is that existence of such an estimator would allow one to distinguish between violations of A3 and A4 from the data alone, which in turn would imply the ability to learn about the unobserved confounder-induced selection bias, which we cannot characterize with the observed data by definition. Given that a doubly resilient estimator does not exist, we now continue with the exposition of our proposed estimator. 

\section{Falsification Test for External Validity and Conditional Ignorability}\label{sec:testing}
As stated initially, we would like to leverage the experimental sample in order to learn something about the observational sample, and vice versa. Here, we show that it is possible to test whether at least one of the assumptions A3 and/or A4 are violated, without necessarily knowing which assumption(s) are violated. This is done by demonstrating that the same test statistic provides us evidence against the null of A3 and A4 holding towards the following two alternatives: 
\begin{enumerate}
\item Given that internal validity (A2) and external validity (A3) of the experiment hold, we can test for violations of ignorability in the observational sample (A4),  
\item Alternatively, given that internal validity of the experiment (A2), and ignorability of treatment (A4), we can test for violations of external validity in the experimental sample (A3).
 \end{enumerate}

To formalize this claim, we introduce the quantity $\theta(t) := \mathbb{E}_{\bX}\left[ \mu(\bX,t,0) - \mu(\bX,t,1) \right]$ that can provide insight into the existence of an unobserved confounder. Our key claims 
are as follows:

\begin{theorem}[Identification]\label{thm:id}~
\begin{enumerate}
    \item Given the set of assumptions (A1, A2, A3), if there exists a $t$ such that $\theta(t) \neq 0$, then A4 (conditional ignorability in the observational sample) is violated.
    \item Given the set of assumptions (A1, A2, A4, A5), if there exists a $t$ such that $\theta(t) \neq 0$, then A3 (external validity in the experimental sample) is violated.
    \item $\E[Y(t)|\bX=\bx]$ is identified under either (A1, A2, A3) or (A1, A2, A4, A5).
\end{enumerate}
\end{theorem}
The proof of this theorem is in Appendix~\ref{sec:proofs}.

The above result implies that $\theta(t)$ is informative about the level of ``confoundedness'' of the data with respect to potential outcome $Y(t)$.  To see this, we can expand the definition of $\theta(t)$ and use Equation~\eqref{eq:6}:
\begin{equation*}
    \theta(t) = \int_{\cX}\int_{\cY}\color{black}y\left[f_{y|\bX=\bx, T=t, S=0}(y) - f_{y|\bX=\bx,T=t,S=1}(y)\right]dy\, dF_\bx.
\end{equation*}
Therefore, $\theta(t)$ will be informative as to the expected difference between the distributions of the potential outcomes across the two samples. When this difference is large in expectation, then $\theta(t)$ will be large (in magnitude), and we can conclude that either sample may not be very informative as to the true value of the mean potential outcome of interest. This may be the case if the selection of units in the experimental sample or the treatment selection in the observational sample are affected by unobserved confounders. Thus, if an analyst believes external validity holds for their data, a large $\theta(t)$ would then suggest violation of conditional ignorability, and vice versa. 




\subsection{An Estimator for Confounding}
\begin{lemma}\label{lemma:test_eif}
    Given assumption A1, A2 and A5, for any $t \in \{0,1\}$, the efficient influence function for $\theta(t)$ under null is given as 
    \begin{eqnarray}
\phi(t) &:=& \mutobs - \mutexp \nonumber + \frac{T(t) (1-S)}{\etobs(1-\pr)}(Y - \mutobs) 
\\&& \quad\quad\quad\quad\quad\quad\quad\quad\quad 
- \frac{T(t)(S)}{\etexp\pr}(Y - \mutexp). \label{eq:psi}
\end{eqnarray}
\end{lemma}
The efficient influence in Lemma~\ref{lemma:test_eif} can be derived via procedure described in \cite{hines2022demystifying} and \cite{kennedy2024semiparametric}.
We now introduce a flexible two-stage estimator for $\theta(t)$ based on the efficient influence function in equation~\ref{eq:psi}. Our estimator leverages the following equality: $\mathbb{E}[\psi(t)] = \theta(t)$ for doubly robust and efficient estimation.

The asymptotic properties of our estimator can be derived specifically for the estimation problem as a two-stage procedure leveraging results on double ML in \cite{chernozhukov2018double}. The first-stage parameters that our estimator of $\theta(t)$ depends on are estimates of all the quantities defined in Equations \eqref{eq:muobs} -- \eqref{eq:pri}. Our framework estimates these quantities using flexible machine-learning models, leading to the theoretical guarantees for our second-stage estimator. 


\noindent Our proposed procedure is comprised of two stages:\\
\textbf{Stage 1}: Randomly partition the full sample into \(K\) approximately equal-sized folds, indexed by \(k = 1, \dots, K\). For each fold \(k\), proceed as follows:
\begin{enumerate}
    \item Fit a flexible ML model for predicting \(\mu(\bx_i,t,0)\) using the \textbf{observational} units not in fold \(k\). Use the fitted model to obtain predictions \(\hat{\mu}(\bx_i,t,0)\) for units in fold \(k\).
    \item Fit a flexible ML model for predicting \(\mu(\bx_i,t,1)\) using the \textbf{experimental} units not in fold \(k\). Use the fitted model to obtain predictions \(\hat{\mu}(\bx_i,t,1)\) for units in fold \(k\).
    \item Estimate the sampling propensity \(p(\bx_i)\) either using true known propensities or by fitting a flexible ML model to sample indicators \(S_i\) using units not in fold \(k\), and obtain predictions \(\hat{p}(\bx_i)\) for units in fold \(k\).
    \item Fit a flexible ML model for predicting the treatment propensity \(e(\bx_i,t,0)\) using \textbf{observational} units not in fold \(k\), and obtain predictions \(\hat{e}(\bx_i,t,0)\) for units in fold \(k\).
    \item Fit a flexible ML model for predicting the treatment propensity \(e(\bx_i,t,1)\) using \textbf{experimental} units not in fold \(k\) (or use true experimental propensities if known), and obtain predictions \(\hat{e}(\bx_i,t,1)\) for units in fold \(k\).
\end{enumerate}

\noindent This crossfitting structure ensures that the model used to predict each quantity for a unit is trained on data \textbf{excluding} that unit, thus mitigating overfitting bias.

\vspace{0.3cm}

\noindent\textbf{Stage 2}: After obtaining the first-stage estimates for all units via cross fitting, compute the following estimator:
\begin{equation}
    \thetahat(t) = \frac{1}{N}\sum_{i=1}^N \Psiithat, \label{eq:dmlmean}
\end{equation}
where $\Psiithat =$
\begin{equation}
 \muitobshat - \muitexphat + 
\frac{\Tit (1-S_i)}{\eitobshat(1-\prihat)}(Y_i - \muitobshat) - \frac{\Tit(S_i)}{\eitexphat\prihat}(Y_i - \muitexphat). 
\label{eq:psihat}
\end{equation}

\noindent In this second stage, we aggregate the per-unit corrected estimates \(\Psiithat\) across all units to produce a final estimate of \(\theta(t)\). Each unit’s contribution \(\Psiithat\) is a bias-corrected difference between the observational and experimental response function estimates. If a unit belongs to the experimental sample, its observed outcome is used to correct the experimental model estimate; if it belongs to the observational sample, its observed outcome corrects the observational model estimate. These corrections are critical for achieving asymptotic efficiency.

Owing to the specific form of the estimator, and to the cross-fitting procedure at Stage 1 of our algorithm, the asymptotic properties of $\thetahat(t)$ are given in the following theorem: 
\begin{theorem}\label{thm:dml_theta}
Under null such that Assumptions (A1-A6) are satisfied, then:
\begin{align*}
    \sqrt{n}\thetahat(t) \overset{d}{\rightarrow} \mathcal{N}(0, \sigmasq(t)),
\end{align*}
where: $\sigmasq(t) = \E\left[ \phi^2(t) \right]
= \E\left[(\phi(t) - \theta(t))^2\right]$
and the expectation is taken over $\Om$. \\In addition, the na\"ive estimator for the variance is consistent with the first-stage estimates used in place of the true nuisance parameters:
$$\sigmasqhat(t) = \frac{1}{n}\sum_{i=1}^N(\Psiithat - \thetahat(t))^2 \overset{p}{\rightarrow} \sigmasq(t),$$
Finally, it follows that an uniformly valid asymptotic $1-\alpha$ confidence interval is: $$\left[\thetahat(t) \pm \Phi^{-1}(1-\alpha/2)\sqrt{\sigmasqhat(t)/n}\right]$$ where $\Phi$ is CDF of univariate standard normal distribution.
\end{theorem}
This theorem relies on the adaptation of Theorems 3.1 and 3.2, and Corollary 3.1 in \cite{chernozhukov2018double} to our setting. Its proof, which can be found in the appendix, revolves around showing that the score yielding the estimator $\hat{\Psi}$ enjoys properties like unbiasedness, Neyman-orthogonality, and local insensitivity to the values of the nuisance parameters $\mu(\bx, t, s)$, $e(\bx, t, s)$, and $p(\bx)$. The theorem is useful as it permits us to estimate uncertainty around our estimate of $\theta(t)$, and, therefore, the level of confounding present in the observational dataset. It also allows for the construction of falsification tests regarding the presence of unobserved confounding, as we show next. 

Notably, the above algorithm depends on a leave-one-out style procedure at Stage 1. This procedure can be slow for large datasets and complicated ML models for first stage parameters that take a long time to fit to data. Because of this, a sample-splitting procedure akin to K-fold cross validation is a computationally efficient alternative. Specifically, it is possible to split the data into $K$ non-overlapping folds at random, and to then run both Stage 1 and Stage 2 by fitting models on all data but the units in the $k^{th}$ fold, and then predicting on those units. The final output of this procedure will be $K$ estimators of the form in \eqref{eq:dmlmean}, which can be averaged together to obtain a final estimate of $\theta(t)$. 
This procedure of sample-splitting for efficient estimation of semiparametric models, also known as repeated cross-fitting, has been well studied in the literature \citep{schick1986asymptotically, bickel1993efficient, zheng2010asymptotic, ayyagari2010applications, robins2013new,chernozhukov2018double}. \cite{bickel1993efficient} examined estimation of nuisance functions using a vanishingly small portion of the sample; \cite{schick1986asymptotically} extended these ideas by introducing equal two-way sample splitting and discretization of the parameter space; and van der Vaart’s treatment in \citep{bickel1993efficient} also employed such sample splitting and discretization to provide weak conditions for $k$-step estimators based on efficient scores, with the updates estimated on the split samples. Along similar lines, \cite{robins2013new} proposed sample splitting to construct higher-order influence function adjustments in semiparametric estimation; and in the targeted maximum likelihood literature, \cite{zheng2010asymptotic} operationalized sample splitting in the context of $k$-step updating.

\subsection{Using $\thetahat(t)$ as a test statistic}
Our goal is to test if either A3 or A4 is violated. To do this, we note that 
Theorem~\ref{thm:id} implies that if Assumptions A3 and A4 are satisfied (along with A1, A2, A5, A6) then $Y$ and $S$ are mean independent conditional on $X,T$ i.e., $\E[Y|\bX=\bx_i, T=t, S=1] = \E[Y|\bX=\bx_i, T=t, S=0]$. Thus, violation of mean independence implies violation of either A3 or A4.
This suggests the following specification of the null hypothesis of interest:
\begin{align}
    \Hzerot &: \E[Y|\bX=\bx_i, T=t, S=1] = \E[Y|\bX=\bx_i, T=t, S=0] \label{eq:confNull}\\
            &\equiv \muitobs = \muitexp. \nonumber
\end{align}


By definition, under $\Hzerot$:
\begin{align*}
    \theta(t) &= 0, \text{ and }\\
    \sigmasq(t) &= \E\left[ \E[(Y - \mu(\bX,t,0))^2 | \bX, T=t] \left( \frac{(1-S)}{e(\bX,t,0) (1-p(\bX))} - \frac{S}{e(\bX,t,1) p(\bX)} \right)^2 \right].
\end{align*}
This allows us to use the results in Theorem~\ref{thm:dml_theta} to construct a test of this null. The second statement is a direct consequence of expanding $\sigmasq(t) = \E[(\Psiit - \theta(t))^2|\Hzerot]$ and $\muitobs = \muitexp$ under $\Hzerot$.  
%
First we note that under the assumptions of Theorem \ref{thm:dml_theta}, a consistent estimator for the variance above is given by: 
\begin{equation}
\sigmasqhat_r(t) = \frac{1}{n}\sum_{i=1}^n\left( \left(\frac{(1-S_i)\Tit(Y_i - \muitobshat)}{\eitobshat(1-\prihat)}\right)^2 + \left(\frac{S_i\Tit(Y_i - \muitobshat)}{\eitexphat\prihat}\right)^2 \right).\label{eq:sigmahatrest}
\end{equation}
Since, under $\Hzerot$, the required assumptions for the theorem are satisfied, this allows us to construct a Wald-type test statistic $\thetahat(t)$: $Z_n(t) := \sqrt{n}\thetahat(t)/\sigmasqhat_r(t)$. By Theorem \ref{thm:dml_theta}, we know that $Z_n(t)$ is asymptotically normal under $\Hzerot$. By construction, it is enough to reject the null for a single $t$ in order to have evidence against both A3 and A4 holding. In some settings only a single test need to be performed (Section~\ref{sec:lalonde}) but when multiple tests are performed, a correction for multiple falsification testing, such as Bonferroni's correction, may become needed.

\section{Treatment Effect Estimation}\label{sec:drate}
In the presence of an unobserved confounder, it is not possible to consistently estimate the ATE using just observational data without any additional untestable assumptions. This section presents a regular and asymptotically normal estimator that combines experimental and observational data to consistently estimate the ATE when \textit{conditional ignorability (A4) is violated} but \textit{external validity (A3) holds.}

\subsection{Efficiency Bound \& Asymptotic Variance}
We start by first deriving an efficient influence function (EIF) in a scenario where (i) both A3 and A4 hold and (ii) A3 holds but A4 is violated (the scenario mentioned above). 

The EIF is the canonical gradient of the pathwise derivative of the statistical estimand, capturing the most informative direction for local perturbations of the data distribution within the nonparametric model \citep{rudolph2023improving}. As such, is used as a foundation for deriving the efficient estimator with the lowest possible asymptotic variance, as defined by the efficiency bound. This bound represents the smallest achievable variance for an unbiased regular and asymptotically linear (RAL) estimator under the given model constraints. We use this EIF to guide the construction of a robust double machine learning estimator. 

The estimand of interest is the average treatment effect (ATE) $\tau:= \E_{\Omega}[Y(1) - Y(0)]$. For simplicity of notation, let
$\nu(t) := \E[Y(t)]$ be the expected potential outcome for treatment $t$ and $\nu(\bx,t) := \E[Y(t) \mid \bX=\bx]$ be the conditional equivalent of the same. Then, the ATE can be rewritten $\tau = \nu(1) - \nu(0)$. Further, let  $\forall \bx,t,s, Var(Y \mid \bX=\bx, T=t, S=s) = Var(Y \mid \bX=\bx, T=t)$ (cross-sample homoskedasticity) and $\sigma^2_Y(\bX,t) = Var(Y \mid \bX,T=t)$.
We use the strategies delineated in \cite{hines2022demystifying} and \cite{kennedy2024semiparametric} to derive the efficient influence function for $\nu(1)$ and $\nu(0)$ and the corresponding efficiency bound.

\begin{theorem}[EIF under A3]\label{th:eif_a3}
Given assumption A1-A3 and cross-sample homoskedasticity for any $\bx, t$
    \begin{enumerate}
        \item If A4 holds then the efficient influence function for estimating $\nu(t)$ is given as 
        \begin{eqnarray*}
            EIF(t) &=& \mutexp \pr + \mutobs (1-\pr) - \nu(t) \\&&
                    + \left( \frac{S \mathbf{1}[T=t]}{\etexp}(Y - \mutexp) + \frac{(1-S) \mathbf{1}[T=t]}{\etobs}(Y - \mutobs)\right) 
        \end{eqnarray*}
        and the efficiency bound is given as follows 
        \begin{eqnarray*}
             \E\left[ \frac{\sigma^2_Y(\bX,t) p(\bX) }{e(\bX,t,1)} + \frac{\sigma^2_Y(\bX,t) (1-p(\bX))}{e(\bX,t,0)} \right]+ \E\left[ (\nu(\bX,t) - \nu(t))^2\right] 
        \end{eqnarray*}
        \item If A4 is violated then the efficient influence function for estimating $\nu(t)$ is given as 
        \begin{equation*}
            EIF(t) = \mutexp + \frac{S}{\pr}\left( \frac{\mathbf{1}[T=t]}{\etexp}(Y - \mutexp)\right) - \nu(t)
        \end{equation*}
        and the efficiency bound is given as follows 
        \begin{equation*}
             \E\left[\frac{\sigma^2_Y(\bX,t)}{p(\bX)e(\bX,t,1)} \right] + \E\left[ (\nu(\bX,t) - \nu(t))^2\right] 
        \end{equation*} 
    \end{enumerate}
    
\end{theorem}

Theorem~\ref{th:eif_a3} characterizes the efficiency bounds, i.e., the smallest achievable variance for unbiased, RAL estimators, under two distinct sets of assumptions. Comparing the efficiency bounds in Theorems~\ref{th:eif_a3}.1 and \ref{th:eif_a3}.2, we find that the efficiency bound under assumption A3 alone (i.e., violation of A4) is larger than under both assumptions A3 and A4 -- this is as expected. Furthermore, under assumption A3 without A4, the EIF utilizes treatment-effect information exclusively from the experimental data and does not leverage outcome or treatment data from the observational study. This highlights that when conditional ignorability (A4) is violated, \textit{no asymptotic efficiency gain} is achievable by incorporating observational data. There are no RAL estimators of any marginal treatment effect that achieve a lower efficiency bound than the AIPW estimator using data only from the RCT under assumption A3 (without assumption A4). Nevertheless, even in the absence of asymptotic benefits, observational study data may still offer finite-sample variance improvements (albeit at the cost of finite-sample bias), particularly if the observational dataset is considerably larger than the experimental one. In the next section, we introduce an estimator that effectively incorporates observational study data (under the violation of A4), which may result in finite-sample benefits.

\subsection{Estimation under External Validity (A3)}
We build on the insights from Theorem~\ref{th:eif_a3} in order to propose an estimator for the ATE that incorporates both observational and experimental data. 
Consider $\Lambdait =  \muitobs + \frac{S_i \Tit}{\pri \eitexp}(Y_i - \muitobs )$, then under assumptions A1-A3 $\nu(t) = \mathbb{E}[\Lambdait]$.
We first estimate $\nu(t)$ by estimating $\Lambdait$ for each $i$ and $t$, and then estimate the ATE as the difference of $\nu(1)$ and $\nu(0)$.  We construct our cross-fitted estimator for $\nu(t)$ as:

\begin{equation}
    \nuhat(t) = \frac{1}{n}\sum_{i=1}^n \Lambdaithat, \label{eq:nuhatdml}
\end{equation}
where:
\begin{equation}
\Lambdaithat =  \muitobshat 
+ \frac{S_i \Tit 
(Y_i - \muitobshat 
)
}{\prihat \eitexphat}. \label{eq:lambdahat}
\end{equation} 

This estimator is structured around the doubly-robust estimator studied in, e.g., \citet{glynn2010introduction, farrell2015robust}. First, conditional means and propensities are estimated for all units (both experimental and observational) by fitting flexible non-parametric ML models. 
Second, these estimates are combined for each unit ($\Lambdaithat$), and the scores are then averaged to obtain an estimate $\nuhat(t)$. An estimate of $\tau$ can be obtained simply by taking the difference of $\nuhat(1)$ and $\nuhat(0)$. 

In our case, the estimator we propose is consistent whenever external validity (A3) is satisfied, even if conditional ignorability (A4) is violated.
We provide a formal derivation in the supplement. 

On top of this guarantee of consistency under violation of A4, the results for two-step estimators of \citet{chernozhukov2018double} apply to our estimator for $\nu(t)$, and permit us to state the following properties of our method:
\begin{theorem}\label{thm:dml_pate} 
Given assumptions A1 -- A3 are satisfied, if $\hat{p}(\bX) \rightarrow p(\bX) $ and $\hat{e}(\bX,t,1) \rightarrow e(\bX,t,1) $ then $\hat{\nu}(t)$ is a consistent estimator of $\nu(t)$ i.e. $\hat{\nu}(t) \overset{p}\rightarrow \nu(t).$
Further, if Assumptions A4 and A6 are also satisfied, then $\sqrt{n}(\nuhat(t) - \nu(t)) \overset{d}{\rightarrow} \mathcal{N}(0, \gammasq(t)),$ where $\gammasq(t) = \sigma^2_Y(t) \E\left[\frac{1}{p(\bX)e(\bX,t,1)} \right] + \E\left[ (\nu(\bX,t) - \nu(t))^2\right]$.
In addition, the vanilla estimator for the variance is consistent with the first-stage estimates used in place of the true nuisance parameters,
$\gammasqhat(t) = \frac{1}{n}\sum_{i=1}^N(\Lambdaithat - \nuhat(t))^2 \overset{p}{\rightarrow} \gammasq(t),$
and it follows that an uniformly valid asymptotic $1-\alpha$ confidence interval is $\left[\nuhat(t) \pm \Phi^{-1}(1-\alpha/2)\sqrt{\gammasqhat(t)/n}\right].$
\end{theorem}

The data-splitting procedure at step 1 of our algorithm permits our second-stage estimator to have an asymptotically normal distribution, with known variance. The theorem directly implies the following corollary for the observational ATE: 
\begin{corollary}
Let all the conditions in Theorem \ref{thm:dml_pate} hold, with $\etab$ and $\etabhat$ denoting estimands and respective estimators for two treatment levels, $t$ and $t'$. Then the estimator: $\hat{\tau} = \frac{1}{n}\sum_{i=1}^n\widehat{\Lambda}_i(1) - \widehat{\Lambda}_i(0)$ satisfies: $\sqrt{n}(\hat{\tau} - \tau) \overset{d}{\rightarrow}\mathcal{N}(0, \Gamma^2)$, with $\Gamma^2 = \E[(\Lambda_i(1) - \Lambda_i(0) - \tau)^2]$. Additionally, the estimator $\widehat{\Gamma}^2 = \frac{1}{n}\sum_{i=1}^n(\widehat{\Lambda}_i(1) - \widehat{\Lambda}_i(0) - \hat{\tau})^2$ is consistent for $\Gamma^2$. 
\end{corollary}
The usefulness of this corollary is in that it permits us to construct approximate hypothesis tests and confidence intervals for $\tau$, with guaranteed uniform asymptotic coverage. 

We provide the proof of Theorem~\ref{thm:dml_pate} in Appendix~\ref{sec:proof_thm_dml_pate}. As for Theorem \ref{thm:dml_theta}, the proofs rely on showing that the score that has $\hat{\Lambda}$ as an estimator satisfies several key properties, including unbiasedness, Neyman-orthogonality, and local insensitivity to the values of the nuisance parameters. 

\section{Student Teacher Achievement Ratio (STAR) Project}\label{sec:star}
\subsection{Data Description}
Project STAR (Student-Teacher Achievement Ratio) was a three-phase experiment designed to study the effect of class-size on short and long-term student performance \citep{project_star_data,mosteller2014tennessee}. The project was run across 79 schools in the state of Tennessee across inner-city (17), urban (8), sub-urban (16), and rural (38) areas. A single cohort of students was studied within each school for four years. Within each STAR school, students in the study cohort were randomly assigned to one of three treatment arms: a small class (13 to 17 students), a regular class (22 to 25 students), or a regular class with a full-time teacher aide. Teachers were randomized across the treatment arms. The treatment arm was fixed for students who moved between STAR schools during the duration of the study. However, some students participated in the study for fewer than four years if they moved to a non-STAR study school. The student's gender, race, birth year, birth month, and free-lunch status were collected to look for any systematic difference between treatment arms. The primary study did not find any evidence of systematic differences across treatment arms on observable characteristics --- providing evidence against systematic failures in the randomization process, though balance on unobservables cannot be directly verified. Student performance was measured once a year, from 1986 to 1989, using standardized achievement tests, which included norm-references and criterion-referenced tests. Norm-referenced tests included Stanford achievement tests (SAT) which were developed by the Psychological Corporation. These tests are composed of separate reading, mathematics, and listening portions for grades K through 3. The observational comparison group for the study included 1780 students across grades 1 to 3 from 21 schools which were located within the same 13 districts as STAR schools. This was done to ensure a level of similarity between the comparison schools and the STAR schools in their respective districts. The same achievement tests were administered in 1987, 1988, and 1989.

\subsection{Analysis and Result}
Here, we study whether the Project STAR observational data has an unobserved confounder that affects students' selection into classrooms of different sizes. We also assess the degree of the associated selection bias and estimate the average treatment effect of class size on short-term standard test outcomes. Given that comparison, schools were chosen from the same 13 districts in Tennessee as STAR schools and that their similarity to STAR schools was built into the study design, it is reasonable to assume that the external validity of the experiment holds. Our framework thus allows us to test for violation of A4 and estimate the ATE even in the presence of unobserved confounding.

\begin{figure}
    \centering
    \includegraphics[width=0.7\textwidth]{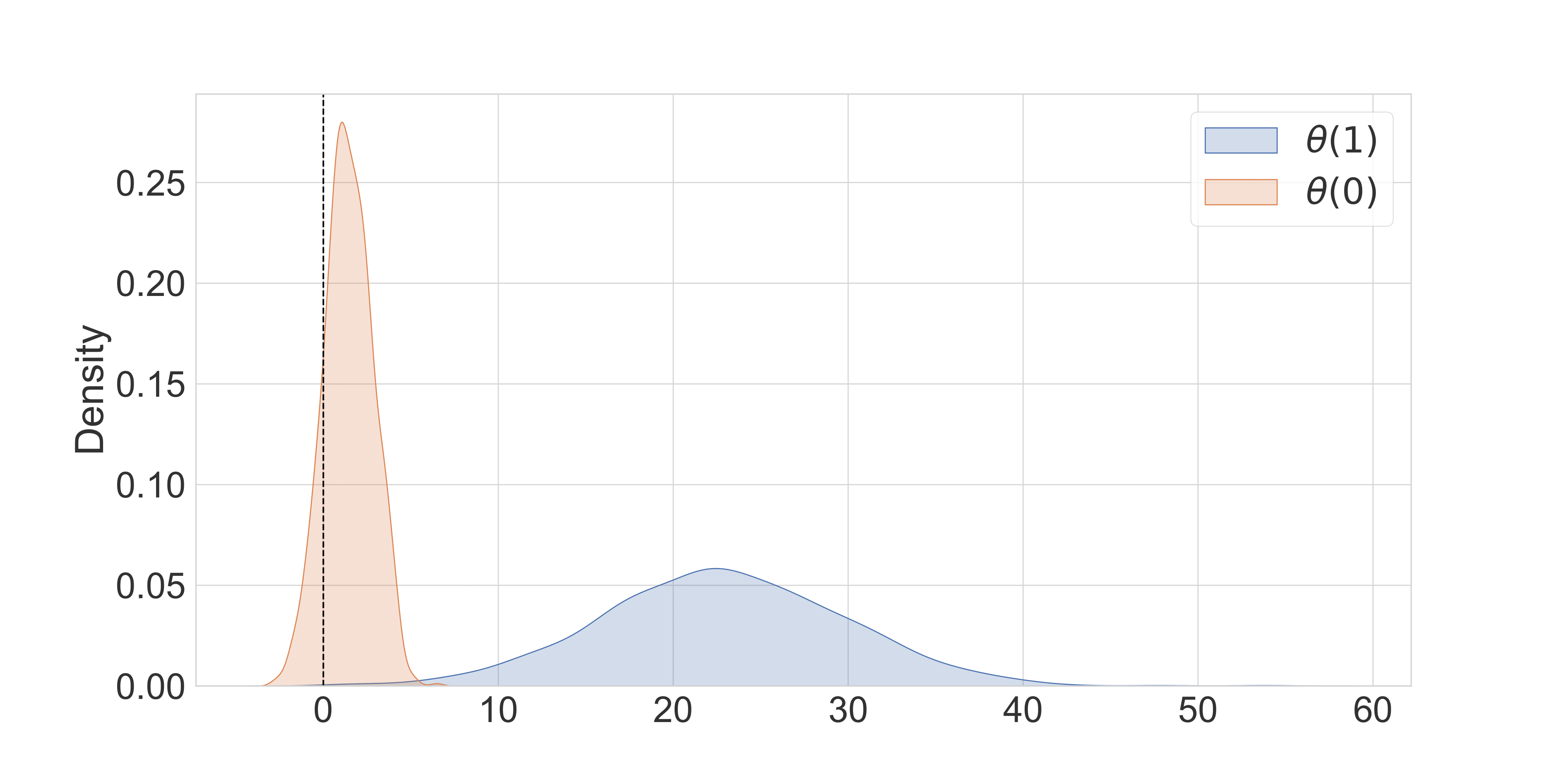}
    \caption{Kernel Density Estimate of the distribution of $\hat{\theta}(1)$ and $\hat{\theta}(0)$. The majority of mass for each of the density functions is in the $>0$ region with minimal density around 0. }
    \label{fig:theta_star}
\end{figure}

\begin{figure}
    \centering
    \includegraphics[width=0.7\textwidth]{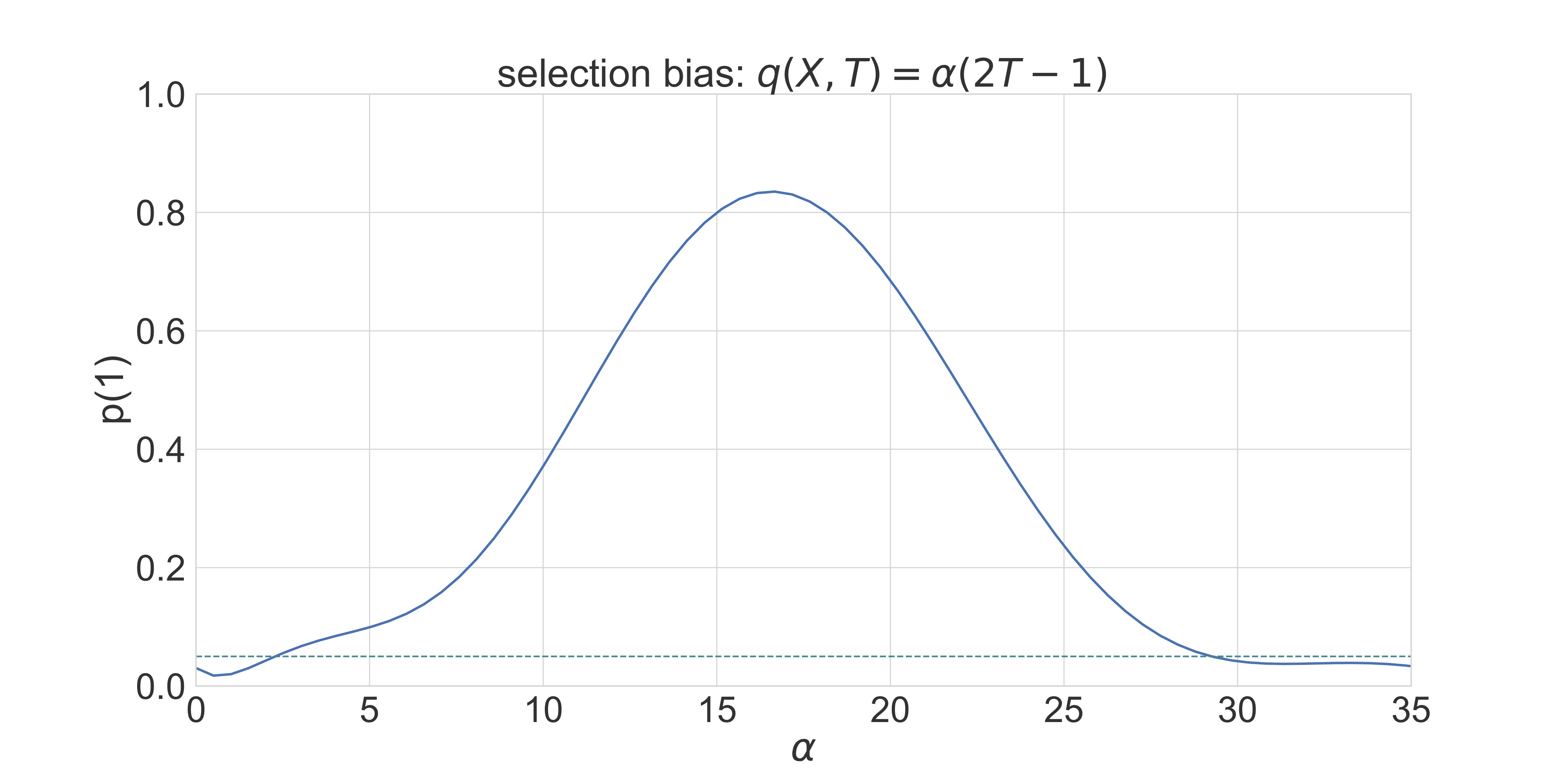}
    \caption{Confounding analysis to estimate the level of bias affecting the selection into small classrooms. Potential estimates of level of selection bias are all values of $\alpha$'s for which the test fails to reject the null hypothesis. We find that the null hypothesis is not for $3\leq \alpha \leq 29$ and for $\alpha=16$, the p-value $p(1)$ peaks.}
    \label{fig:sens_star}
\end{figure}

Point estimates of the test statistics for $\theta(1)$ and $\theta(0)$ are $22.96$ and $1.42$, and their distribution is plotted in Figure~\ref{fig:theta_star}. P-values for the two statistics are $0.0006$ and $0.2663$, respectively, meaning that the test rejects that $\theta(1) = 0$ at the (Bonferroni corrected) $0.05/2$ level. 
We thus reject the null and, given the plausibility of external validity discussed above, conclude there is strong evidence for an unobserved confounder affecting selection. 
To analyze the strength of selection bias we employ the idea of breakdown-frontiers: we specify a form for the selection bias, adjust our observed data according to that form, and perform our proposed test using the debiased outcomes \citep{masten2020inference}. When the test \textit{does not} reject, we have a plausible magnitude for the selection bias \citep{blackwell2014selection}. The parametric form we choose is $q(X,T;\alpha) = \alpha(2T-1)$ --- that is, the treated potential outcome for treated units is $\alpha$ larger while the control potential outcome for control units is $\alpha$ smaller. 
We concentrate on $\theta(1)$ and find that the test fails to reject the null hypothesis for $\alpha \in [3, 29]$ (see Figure~\ref{fig:sens_star}). 
Specifically, we observe that for $\alpha=16$, the p-value $p(1)$ peaks and it decreases as we further increase $\alpha$ (see Figure~\ref{fig:sens_star}). 
To contextualize these values of $\alpha$, we note that the outcome is bounded between 400 and 800 and so the relative range of selection bias compared to the scale of $Y$ ranges from $0.75\%$ to $7.25\%$. 

Lastly, we estimate the average treatment effect of small class size on grade 3 standardized test scores using the difference of means estimator for experimental data and compare it with our estimator discussed in Section~\ref{sec:drate}. We estimate the ATE to be $5.75$ 
, in agreement with the difference of means estimate from experimental data of $7.24 \pm 2.72$.

\subsection{Data Description}
The coronary artery surgery study (CASS) was initiated by National Heart, Lung and Blood institute (NHLBI) to study the effect of coronary bypass surgery in comparison to conventional medical therapy. The data were collected from 15 medical centers across the US and Canada from 1974 to 1979, yielding 24,989 patients.
Of these, 2,099 eligible patients were selected for the randomized control trial (RCT) and were part of a comprehensive follow-up study. 780 of the 2,099 patients accepted the randomized treatment -- we refer to them as the experimental arm in our analysis. 1,319 patients refused the randomization and self-selected into treatment groups -- this is our observational arm. For our analysis, in the experimental arm, the treatment group is defined based on intent-to-treat by the original randomized assignment. Note that $23.5\%$ of patients assigned to medical therapy had bypass surgery within 5 years since their angina worsened. In the observational arm, a similar treatment group was identified as any patient who was selected for surgery within 90 days of enrollment or if their surgery was in the first year period (when 95\% of CASS experimental arm surgeries were done). Any observational study patient who did not have early elective surgery were treated as the control group (or medical therapy arm). Here, we use all-cause mortality during the course of the study as the outcome of interest.

\subsection{Analysis and Result}
In this paper, we are interested in studying if the conditional ignorability of the observational sample as well as the external validity of the experimental sample holds. Furthermore, we are also interested in estimating the population average treatment effect of surgical intervention as compared to medical therapy. 

Given that the same set of practitioners administers the treatments for both the observational as well as experimental samples, it is reasonable to assume that selection in the experimental (or observational) arm does not have a direct causal link to the outcome. That implies that Assumption A5 is likely to hold. 

Our test allows us to study if either A3 or/and A4 are violated. If we find significant evidence for $\theta(t)\neq0$ for any $t\in\{0,1\}$, then either of these assumptions can be violated. The result is that our test \textit{fails to reject the null hypothesis} and finds \textit{no  evidence} for the violation of these assumptions. The p-values corresponding to the tests are 0.290 and 0.915, respectively for $\theta(0)$ and $\theta(1)$. We also show the distributions of $\theta(0)$ and $\theta(1)$ in Figure~\ref{fig:theta_cass}.
\begin{figure}
    \centering
    \includegraphics[width=0.6\textwidth]{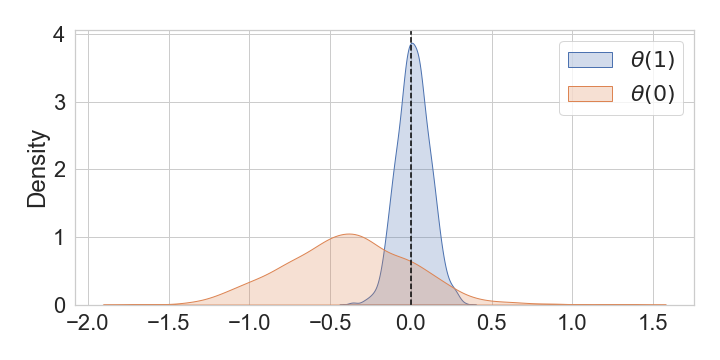}
    \caption{Kernel density plot for $\theta(0)$ and $\theta(1)$ for the CASS dataset. }
    \label{fig:theta_cass}
\end{figure}
Note that failure to find evidence does not necessarily imply that there is no unobserved confounding. However, these results are in congruence with previous works \citep{dahabreh2020extending,olschewski1992analysis}, suggesting that the experimental sample had external validity and the observational sample has conditional ignorability.

Next, we estimate the average treatment effect using our estimator and other estimators. Table~\ref{tab:cass_ate} presents the estimated average treatment effects using our estimator proposed in Equation~\eqref{eq:nuhatdml},  the difference in means estimator for the experimental sample and an augmented inverse propensity score weighted (AIPW) estimator in the observational sample. Since our tests failed to reject, it is not surprising that the three estimates are effectively the same and consistent with the literature: the treatment effect is not significantly different than 0, i.e., the coronary bypass surgery neither helps nor hurts patients' chances of survival. 

\begin{table}
    \caption{\label{tab:cass_ate} ATE Estimates CASS data using different approaches on experimental and observational samples.}
    \centering
    \begin{tabular}{lrr}
    \toprule
    {} &    \textbf{Estimated ATE} &    \textbf{Std. Error} \\
    \midrule
    Our ATE Est.       & -0.019 &  0.024 \\
    Exp. ATE Est.      & -0.013 &  0.032 \\
    Obs. ATE Est.      & -0.022 &  0.034 \\
    \bottomrule
    \end{tabular}
\end{table}
\subsection{Data Description}
The Lalonde data consist of a randomized control trial, the National Support Work Demonstration (NSW), that studied the effect of training programs on participant income levels \citep{lalonde1986}. The NSW is often augmented with the Population Survey of Income Dynamics (PSID-2) dataset of observational control units \citep{dehejia1999}; together, the two can serve as a benchmark for observational causal inference methods. While most estimation methods struggle with recovering the experimental point estimate using the joint dataset, some methods are able to recover the experimental ATE after pre-processing \citep[e.g.,][]{malts}. We reanalyze the NSW and PSID-2 datasets to study if the experimental controls are comparable to observational controls; that is, if the experiment satisfies external validity (Assumption A3).
We limit our analysis to the subsample of male household heads under the age of 55 and who are not retired by 1975. The outcome of interest is the income of participants in 1978. Furthermore, for both datasets, we use pre-treatment information about units' age, race, marital status, education, and income in 1975. 

\subsection{Analysis and Results}
Unlike our analyses of the STAR and CASS datasets, the observational data in this analysis consists only of control units. Our analysis is thus limited to identifying a possible violation of experimental external validity (A3) by testing if $\theta(0)=0$. Our test rejects the null hypothesis (p-value = $6.2 \times 10^{-6}$), indicating that the NSW experiment lacks external validity (with respect to PSID-2). As such, it is not at all surprising that many observational causal inference approaches are unable to recover experimental ATE when using the observational control group. One notable exception to this behavior is using the matching algorithm MALTS \citep{malts}.

We reanalyze the joint NSW and PSID-2 data using MALTS, a matching algorithm that learns a distance metric to guarantee tighter matches on more important covariates. In this approach, we estimate a matched group of size 10 for each unit in the sample and calculate a diameter of the matched group with respect to the learned metric. We plot these diameters in Figure~\ref{fig:prune_lalonde} and note the large gap between diameters of size 80 and 100. Using this heuristic we prune (eliminate) units from the analysis that have a diameter that's larger than 80; we emphasize that we do \textit{not} use outcome information in order to prune.
It is important to note that the pruned set has a significant number of control units from the observational arm along with units from the experimental arm as shown in Figure~\ref{fig:prune_lalonde} by the color of the marker. 

After pruning, the test for violation of A3 then \textit{fails} to reject the null (p-value = $0.79$ and Figure~\ref{fig:theta_lalonde} shows the point estimate of $\theta_0$ with respect to the reference null distribution before and after pruning). While the matched groups were pruned only based on the tightness of the match, the corresponding change in the test statistic suggests that it is possible that an  unobserved confounder causing selection in the experiment is actually \textit{correlated} with observed pre-treatment covariates. This provides a compelling explanation for why most observational approaches fail to recover the experimental ATE using an observational arm, while a flexible matching framework is able to do so. 

\begin{figure}
    \centering
    \includegraphics[width=0.6\textwidth]{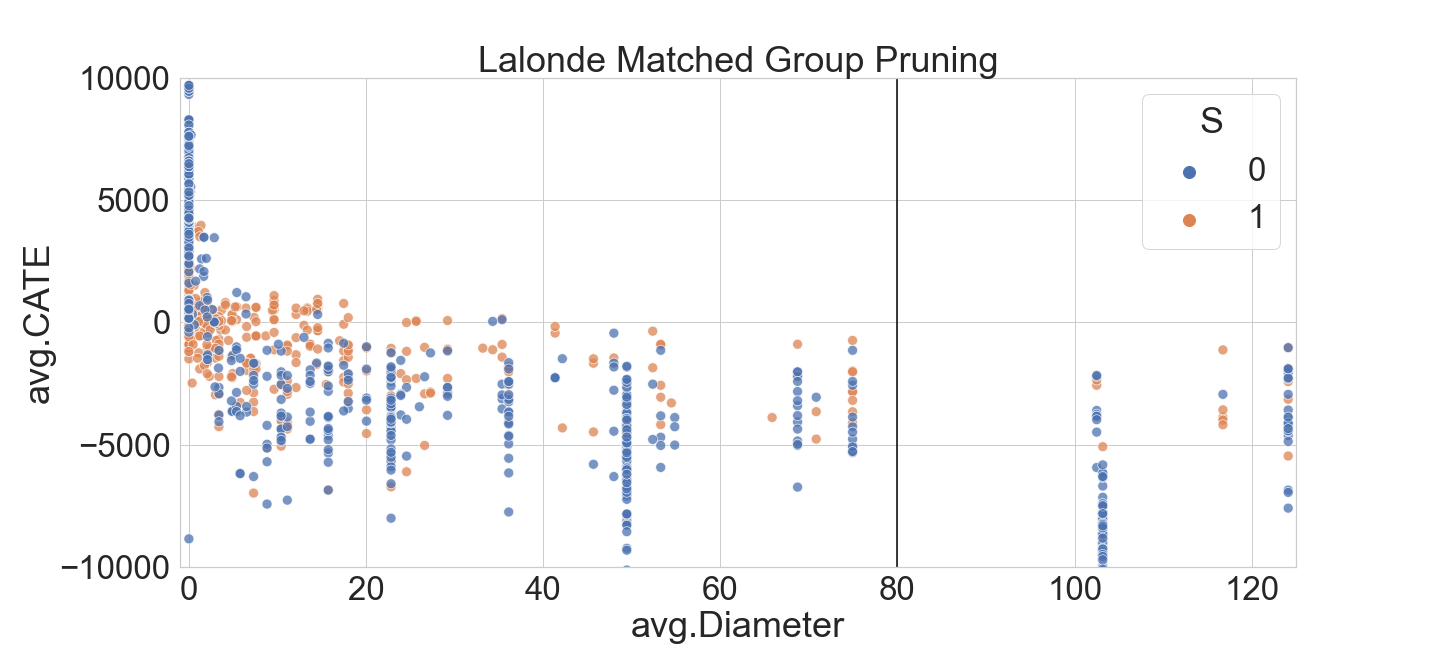}
    \caption{Pruning criteria for matched groups created by Lalonde based on the tightness of matches (measured as the diameter of the matched group). The black vertical line at 80 denotes the threshold above which matched groups are eliminated because the control and treated units are no longer well-matched.}
    \label{fig:prune_lalonde}
\end{figure}

\begin{figure}
    \centering
    \includegraphics[width=0.6\textwidth]{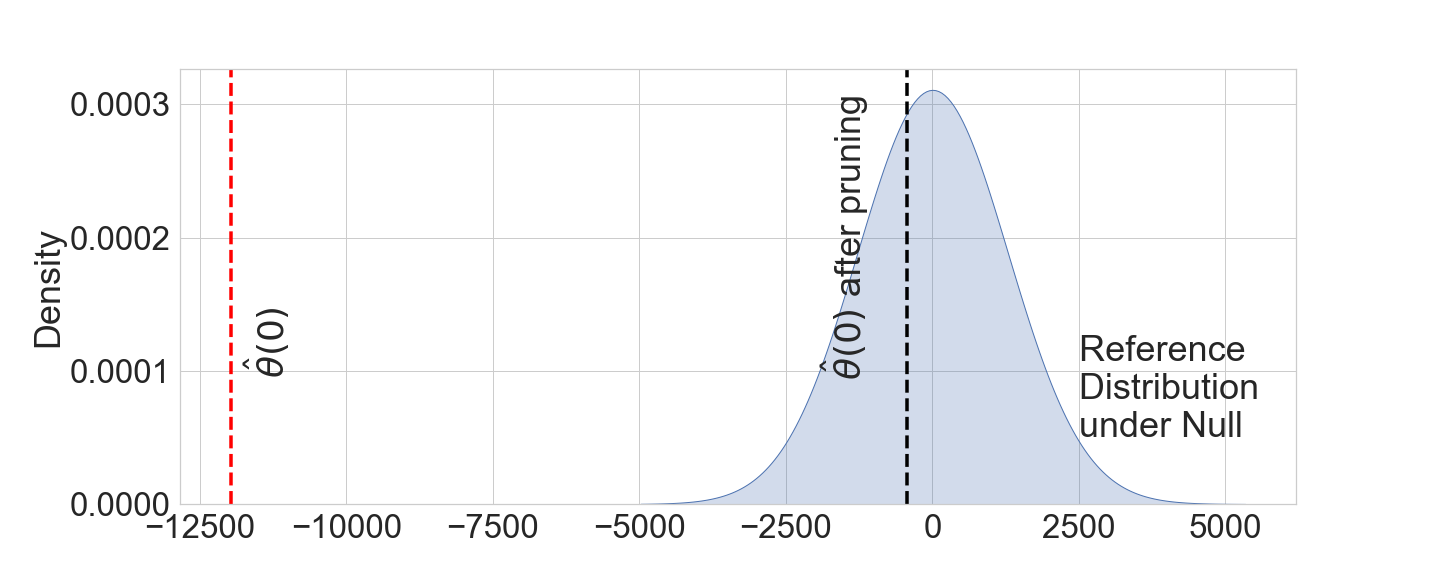}
    \caption{Distribution of the null distribution (blue) and point estimate of $\hat{\theta}(0)$ (red) in the Lalonde data.}
    \label{fig:theta_lalonde}
\end{figure}
\section{Synthetic Data Experiment} \label{sec:synth_exp}
In this section, we contrast our Double Machine Learning (DML) approach with the state-of-the-art data fusion methods that similarly aim to blend experimental and observational data for a more precise estimation of treatment effects, employing a straightforward yet illustrative synthetic data experiment. We begin by detailing the process of generating our synthetic data and describing the alternative baseline approaches. This is followed by an analytical comparison of these methodologies, highlighting how our DML strategy enhances the integration of insights from both experimental and observational studies.

\paragraph{Data Generative Procedure.} We generate the data that satisfies assumptions A1-A3 and A4 is violated. We consider the dataset with 3 pre-treatment covariates $X_0 \dots X_3$ and an unobserved covariate $U$ sampled identically and independently from the normal distribution $\mathcal{N}(1/2, 25)$. The study participation, corresponding treatment, and outcomes are determined as follows:
\begin{eqnarray*}
    S &\sim& \text{Bernoulli}(0.5 \times \text{expit}(X_1 - X_2)) \\
    T &\sim& 
        \begin{cases} 
        \text{Bernoulli}(0.5) & \text{if } S = 1 \\
        \text{Bernoulli}(\text{expit}(X_1 + X_2 - 2U)) & \text{if } S = 0
        \end{cases} \\
    Y &\sim&  \mathcal{N}(X_0 + 5 X_2 + T (X_1 + U), 1).
\end{eqnarray*}
Note that, here, $X_1$ and $U$ are effect modifiers, and $X_1$ and $X_2$ is a confounder affecting the study participation $S$, treatment choice $T$, and outcome $Y$. 

\paragraph{Baseline Approaches.} Here, we compared the performance of our double machine learning approach in estimating ATE with the 3 other baseline approaches: (i) augmented inverse probability weighted estimator just using the experimental data \citep{chernozhukov2018double}, (ii) comprehensive cohort studies (CCS) described in \cite{lu2019causal}, and (iii) integrative R-learner of heterogeneous treatment effects combining experimental and observational studies (integrative HTE) described in \cite{wu2022integrative}.

\begin{figure}
    \centering
    \begin{tabular}{cc}
        \includegraphics[width=0.49\textwidth]{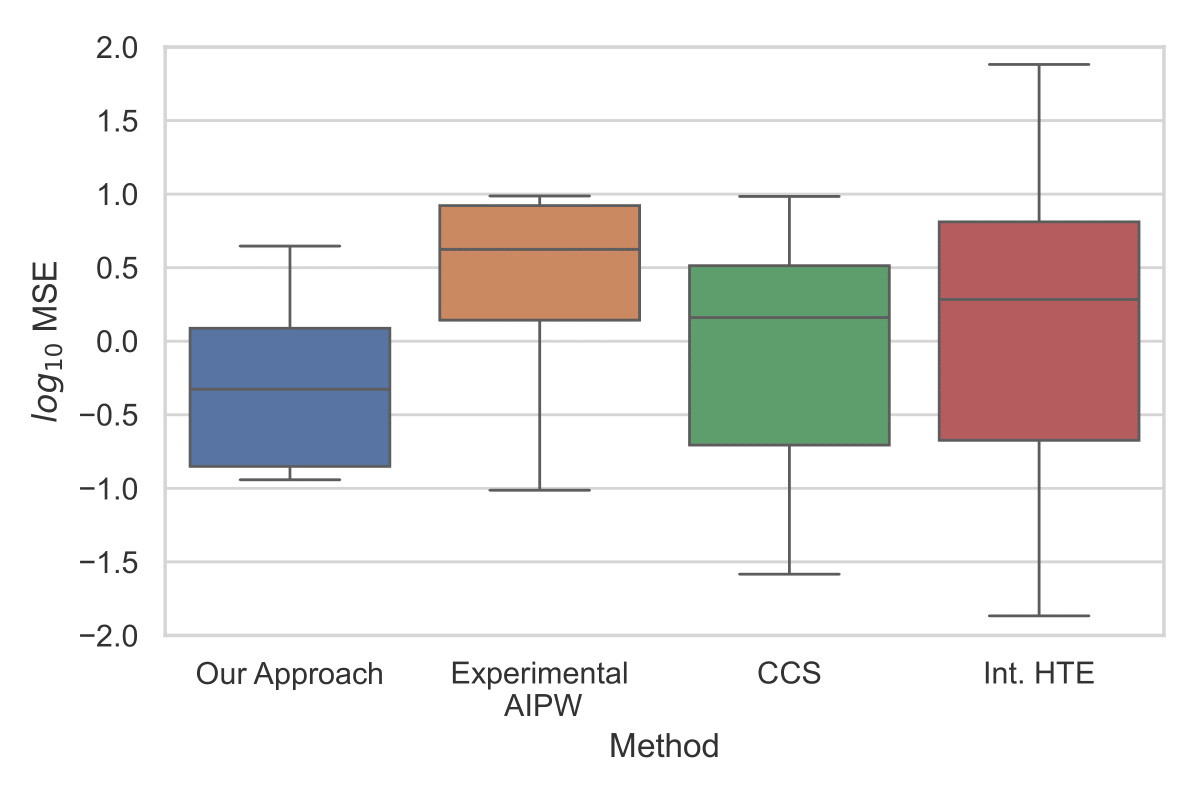} & \includegraphics[width=0.49\textwidth]{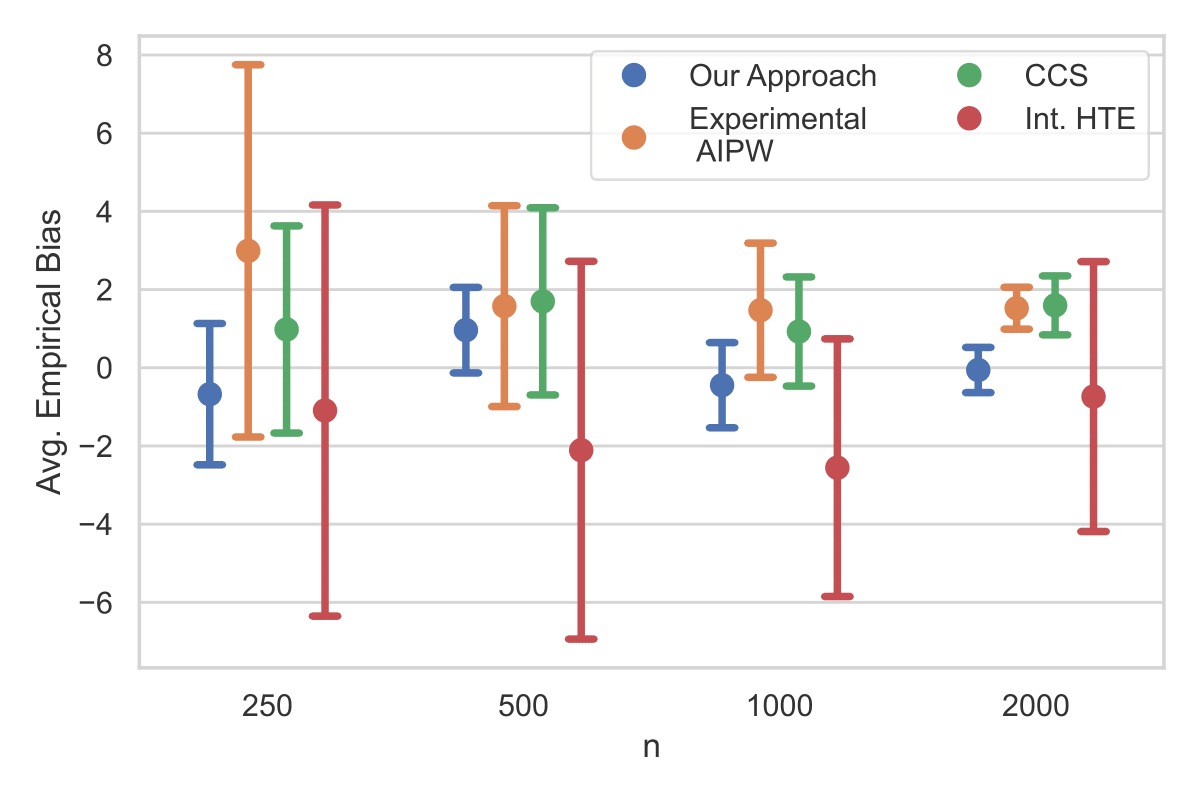}  \\
        (a) & (b)
    \end{tabular}
    \caption{We compare our approach and each of the three baselines approaches over 10 repeats. (a) Mean Squared Error for n=2000 (b) Empirical bias over 10 repeats as the number of samples increases from 250 to 2000. }
    \label{fig:synth_exp}
\end{figure}

\paragraph{Analysis.} We generate datasets of varying sample sizes from $n=250$ to $n=2000$. For the datasets with $n=2000$, our result in Figure~\ref{fig:synth_exp}(a) shows that, on average, our DML approach has the smallest mean squared error of all the other approaches. Further, while comparing the empirical bias and the corresponding variance, Figure~\ref{fig:synth_exp}(b) shows that our DML approach has the smallest spread of empirical bias centered around zero, corresponding to the tight estimates with the smallest standard errors.

\section{Conclusion}\label{sec:conclusion}
With the expanding role of causal inference in all avenues of high-stakes decision-making, it is of fundamental importance that studies making causal claims are as robust to violations of critical assumptions as possible. Experimental studies, such as randomized trials, do not suffer from violations of ignorability assumptions but suffer from small samples and a lack of generalizability to larger populations of interest. On the other hand, while observational studies often involve larger and more representative samples, they are prone to violations of ignorability. 

In this paper, we introduce methods that enable analysts to leverage both experimental and observational data simultaneously, allowing them to detect and correct violations of these assumptions. To detect the violations of ignorability in observational data or external validity in experimental data, we have proposed a statistical quantity that summarizes the extent of violations of ignorability, together with a root-n consistent estimator and falsification test for this quantity. To remedy the lack of conditional ignorability in observational data, we have proposed an estimator of treatment effects that simultaneously makes use of both the observational as well as experimental data. While our work proposes both a test and an estimator, they serve independent purposes, and we do not recommend users to use the same data for testing and estimation, as they can result in post-selection bias. Users may choose to split the dataset and use one half for testing and the other for estimation to ensure valid inference.

Under externally valid experimental data, our estimator is unbiased and consistent. Our methods take advantage of modern machine learning tools for outcome estimation and therefore can achieve an excellent degree of accuracy as shown by the synthetic data experiments. We have shown how our proposed tools can be applied to real-world data by re-analyzing the STAR project, and showing that observational samples are likely biased. We also demonstrate the performance of our approach by reanalyzing the CASS study, showing that the observational sample in this study likely satisfies conditional ignorability (see Appendix~\ref{sec:cass}), and the Lalonde study data where the units in the experimental and observational studies are not exchangeable, in general (see Appendix~\ref{sec:lalonde}).

Natural extensions of our methods include the formulation of alternative test statistics and quantities of interest that describe violations of ignorability, as well as considering methods to assess potential violations of other causal assumptions, such as SUTVA. Ultimately, the methods we have proposed are aimed at strengthening the trustworthiness and reliability of causal inference, as well as enabling analysts to take advantage of all the possible data sources they have access to.

\section*{Data Availability}
The data used to demonstrate the functionality of our methodology is publicly available online. Readers can download (1) Project STAR data from \href{https://dataverse.harvard.edu/dataset.xhtml?persistentId=hdl%3A1902.1%2F10766}{link}, (2) Coronary Artery Surgery Study from \href{https://biolincc.nhlbi.nih.gov/studies/cass/}{link}, and (3) Lalonde NSW and PSID datasets from \href{https://users.nber.org/~rdehejia/data/.nswdata2.html}{link}


\bibliography{biblio}

\newpage

\appendix
\section{Theoretical Results}\label{sec:proofs}
We, first, want to remind readers about that assumptions A1-A6 that will be useful for readers to follow the proofs:

\noindent\textbf{A1 (Data Distribution)} For notation's sake, define the random vector \\$\Om := (Y, T, S, \bX)$. Let $f_{\Om}$ be a probability distribution over the joint domain $\cS = \cY \times \{0, 1\} \times \{0, 1\} \times \cX$. We assume that the full data are iid random samples from this distribution, i.e.: $\{\Om_i\}_{i=1}^n \overset{iid}{\sim} f_{\Om}$, and that $\Om$ (note the absence of indices) denotes an arbitrary draw from $f_{\Om}$. We also maintain the following standard overlap assumption: for all $\om \in \cS$, we assume $0 < f_{\Om}(\om) < 1$. \\
\textbf{A2 (Internal Validity of the Experiment)} We assume that $(Y(0), Y(1)) \indep T \mid \bX=\bx, S = 1$, i.e., if $i$ is an experimental unit, then its treatment assignment is independent of its potential outcomes conditional on its covariates. In other words, $P(Y(t) \mid \bX=\bx, S = 1) = P(Y(t) \mid T=t, \bX=\bx, S = 1)$. 
\\\textbf{A3 (External Validity of the Experiment)} We assume that, adjusted for pre-sampling covariates $\bX$, the units are sampled into the experiment independently of their potential outcomes: $(Y(0), Y(1)) \indep S \mid \bX=\bx$.\\
\textit{As a consequence of A3,} $P(Y(t) \mid \bX=\bx) = P(Y(t) \mid \bX=\bx, S = s)$.\\
\textbf{A4 (Conditional Ignorability)} For all $t$, and almost surely over $f_{\Omega}$ we have $(Y(0), Y(1)) \indep T\mid\bX = \bx, S=0$, i.e., the treatment is assigned independently of potential outcomes in the observational sample.\\ 
\textit{As a consequence of A4,} $P(Y(t) \mid \bX=\bx, S = 0) = P(Y(t) \mid T=t, \bX=\bx, S = 0) = P(Y(t) \mid T=1-t, \bX=\bx, S = 0)$.
\\\textbf{A5 (Sampling Ignorability)} For all $t$ and almost surely over $f_{\Omega}$ we have: $Y \indep S \mid \bX = \bx, T=t, U=u$, i.e., being sampled into the experimental set is independent of the outcome conditionally on the covariates, unobserved confounder, and the treatment level. 
\\\textbf{A6 (Nuisance Parameter Estimation: No Violated Assumptions)}
If Assumptions A1 -- A5 hold, it is possible to also impose the following assumption, for all $t \in \{0, 1\}$ and $s \in \{0, 1\}$:
\begin{gather}
     \ptwo{\hat{\mu}(\bx, t, s) - \mu(\bx, t, s)} \times \ptwo{\hat{e}(\bx, t, s) - e(\bx, t, s)} = o(n^{-1/2}) \label{eq:nuis_est1}\\
     \ptwo{\hat{\mu}(\bx, t, s) - \mu(\bx, t, s)} \times \ptwo{\hat{p}(\bx) - p(\bx)} = o(n^{-1/2})\label{eq:nuis_est2}
\end{gather}
where $\pq{f} = \pq{f(\Omega)} = (\int |f(\omega)|^qd {P}(\omega))^{1/q}$ denotes the $L^q(P)$ norm with respect to the distribution of the data. 

Next, we discuss the theorems and their corresponding proofs.
\setcounter{theorem}{0}
\subsection{Proof of Theorem \ref{thm:impossibility}}
Here, we show that it is impossible to construct a consistent estimator without knowing which one of the assumptions A3 or A4 is violated. We formalize this property in the following definition. 
\begin{definition}[Double Resilience]\label{def: doubleres}Given A1, A2, and A5, a doubly resilient estimator $\gdr[1]-\gdr[0]$ is an unbiased and consistent estimator of conditional average treatment effect if at least one of A3 or A4 is \textit{not} violated. 
More precisely, we say that $\gdr$ is doubly resilient if under assumptions (A1, A2, A3, A5) or under assumptions (A1, A2, A4, A5), then $\forall t,\bX,s$: $\mathbb{E}[\gdr]  = \mathbb{E}\left[ Y(t) \mid \bX \right]$ and $\mathbb{E}[\gdr \mid S=s]  = \mathbb{E}\left[ Y(t) \mid \bX, S=s \right]$.
\end{definition}
In what follows, let A1, A2, and A5 hold and exactly one of A3 or A4 is violated. However, we do not, a priori, know which one of A3 or A4 is violated. One can imagine using the test described in Section~\ref{sec:testing} to check if A3 or A4 is violated -- this test does not, however, reveal which one of them is violated. 

For brevity, it will be useful to define $A^\varnothing$ to be the \textit{unknown} set $A^{\varnothing} \in \{\{A3\}, \{A4\}, \{A3, A4\}\}$ of assumptions that is \textit{not} violated. We emphasize that when we reference $A^\varnothing$, we assume no knowledge of the valid assumptions, only that there is at least one. 

~\\
\noindent The proof of our main result will depend on the following two lemmas. 
\begin{lemma}[Violated Assumptions Imply Selection Bias Unidentifiability]\label{lemma:selection_bias}
Given the assumptions A1, A2, A5, and $A^\varnothing$, the selection bias $\psi(\bX, t) := \mathbb{E}\left[Y(t) \mid \bX, S=0 \right] - \mathbb{E}\left[Y(t) \mid \bX, T=t, S=0 \right]$ is not identifiable.
\end{lemma}
\begin{proof}[Proof of Lemma~\ref{lemma:selection_bias}.]
We prove this lemma using proof-by-contradiction. Let's assume $\psi(\bX,t)$ is identifiable under A1, A2, A5, and A$^\varnothing$. We can decompose $\psi(\bX,t)$ as follows:
\begin{eqnarray*}
\psi(\bX,t) &=& \mathbb{E}\left[Y(t) \mid \bX, S=0 \right] - \mathbb{E}\left[Y(t) \mid \bX, T=t, S=0 \right]\\
&=& \mathbb{E}\left[Y(t)\mid \bX, S=0, T=t \right] P(T=t|\bX,S=0) \\
&& + \mathbb{E}\left[Y(t) \mid \bX, S=0, T=1-t \right] P(T=1-t|\bX,S=0) \\ 
&&- \mathbb{E}\left[Y(t) \mid \bX, S=0, T=t \right] \\
&=& \mathbb{E}\left[Y\mid \bX, S=0, T=t \right] P(T=t|\bX,S=0) \\
&& + \mathbb{E}\left[Y(t) \mid \bX, S=0, T=1-t \right] P(T=1-t|\bX,S=0) \\ 
&&- \mathbb{E}\left[Y \mid \bX, S=0, T=t \right],
\end{eqnarray*}
where the second equality is due to the law of total probability, while the third equality is due to consistency (A1) which allows us to replace $E[Y(t)|\bX,S=0,T=t]$ with $E[Y|\bX,S=0,T=t]$.

Further, we can rewrite the decomposition as:  
\begin{eqnarray*}
\psi(\bX,t) &=& ( \mathbb{E}\left[Y(t) \mid \bX, S=0, T=1-t \right] - \mathbb{E}\left[Y\mid \bX, S=0, T=t \right] ) P(T=1-t|\bX,S=0),
\end{eqnarray*}
by noting that $1-P(T=t|\bX,S=0) = P(T=1-t|\bX,S=0)$.
Given A$^\varnothing$, we know at least one of A3 or A4 is not violated. This implies the following:
\begin{itemize}
    \item If \textit{A4 is not violated}, then the first term is identified as follows: $\mathbb{E}\left[Y(t) \mid \bX, S=0, T=1-t \right] = \mathbb{E}\left[Y(t) \mid \bX, S=0, T=t \right] = \mathbb{E}\left[Y \mid \bX, S=0, T=t \right]$, and thus, $\psi(\bX,t) = 0$. This further means that $E[Y(t)|X,S=0] = E[Y|X,T=t,S=0]$
    \item However, if \textit{A4 is violated but A3 is not}, then  $\mathbb{E}\left[Y(t) \mid \bX, S=0 \right] \overset{A3}{=} \mathbb{E}\left[Y(t) \mid \bX, S=1 \right] \overset{(A1,A2)}{=} \mathbb{E}\left[Y \mid \bX, S=1, T=t \right]$. 
    \\Thus, in this case, $\psi(\bX,t) \overset{\textrm{by defn.}}{=} \mathbb{E}\left[Y(t) \mid \bX, S=0 \right] - \mathbb{E}\left[Y(t) \mid \bX, T=t, S=0 \right] \overset{A1-A3}{=} \mathbb{E}\left[Y \mid \bX, S=1, T=t \right] - \mathbb{E}\left[Y \mid \bX, S=0, T=t \right]$, which may or may not be 0.
\end{itemize}

The critical thing to notice from the two bullet points is that the identification strategy for $E[Y(t)|\bX,S=0]$ changes depending on whether A4 is violated. 
Clearly, if $\psi(\bX,t)$ were always zero, this would not be an interesting problem. This means that we need to know which identification strategy to use for $E[Y(t)|\bX,S=0]$ which will in turn identify $\psi(\bX,t)$. Since we do not know which of A3 or A4 is not violated under A$^\varnothing$, $\psi(\bX,t)$ is not identified under A1, A2, A5 and A$^\varnothing$.
%
%
\end{proof}


\begin{lemma}[Double Resilience Implies Identifiability of Selection Bias]\label{lemma:g_dr}
Given A1, A2, A5 and $A^\varnothing$, if an estimator $\gdr$ is doubly resilient, then  then the selection bias $\psi(\bX, t) := \mathbb{E}\left[Y(t) \mid \bX, S=0 \right] - \mathbb{E}\left[Y(t) \mid \bX, T=t, S=0 \right]$ is identified.
\end{lemma}
\begin{proof}[Proof of Lemma~\ref{lemma:g_dr}.]
Assume there exists a doubly resilient estimator $\gdr$. Consider the following quantity: 
\begin{align*}
    \lambda_0(\bX,t) &:= \E\left[\gdr| S=0 \right] - \mu(\bX, t, 0). 
\end{align*}
Given A1, A2, A5 and $A^\varnothing$, $\E[\gdr | S=0] = \E[Y(t)\mid \bX, S=0]$ (by definition \ref{def: doubleres}). Thus, $\lambda_0(\bX,t) = \mathbb{E}[Y(t) \mid \bX, S=0] - \mathbb{E}[Y(t) \mid \bX, T=t, S=0] = \psi(\bX,t)$, hence, identifying selection bias. 
\end{proof}
\setcounter{theorem}{1}
\begin{theorem}[Doubly Resilient Estimators Do Not Exist]
    There does \textit{not} exist any doubly resilient estimator $\gdr$.
\end{theorem}
\begin{proof}[Proof of Theorem \ref{thm:impossibility}.]
Assume there exists an estimator $\g$ that is doubly resilient. Then, by direct application of Lemma 2, the selection bias $\psi(\bX, t) := \mathbb{E}\left[Y(t) \mid \bX, S=0 \right] - \mathbb{E}\left[Y(t) \mid \bX, T=t, S=0 \right]$ is identifiable, contradicting Lemma 1. Thus, $\g$ cannot be doubly resilient. 
\end{proof}

\subsection{Proof of Theorem \ref{thm:id}}
\begin{theorem}[Identification]~
Consider $\theta(t) = \mathbb{E}_\bX[\mutobs - \mutexp]$. Then,
\begin{enumerate}
    \item Given the set of assumptions (A1, A2, A3, A5), if there exists a $t$ such that $\theta(t) \neq 0$, then A4 (conditional ignorability in the observational sample) is violated.
    \item Given the set of assumptions (A1, A2, A4, A5), if there exists a $t$ such that $\theta(t) \neq 0$, then A3 (external validity in the experimental sample) is violated.
    \item $\E[Y(t)|\bX=\bx]$ is identified under either (A1, A2, A3, A5) or (A1, A2, A4, A5).
\end{enumerate}
\end{theorem}
\textbf{For parts (1) and (2)}, if A1 -- A5 hold, then:
    \begin{align*}
        \mutobs - \mutexp &= \E[Y|\bX=\bx, T = t, S = 0] - \E[Y|\bX=\bx, T=t,  S=1] & \\
        &= \E[Y(t)|\bX=\bx, T = t, S = 0] - \E[Y(t)|\bX=\bx, T=t,  S=1] & (\text{A1})\\
        &= \E[Y(t)|\bX=\bx, S = 0] - \E[Y(t)|\bX=\bx, T=t,  S=1] & (\text{A4})\\
        &= \E[Y(t)|\bX=\bx, S=1] - \E[Y(t)|\bX=\bx, T=t,  S=1] &  (\text{A3})\\
        &= \E[Y(t)|\bX=\bx, T = t, S=1] - \E[Y(t)|\bX=\bx, T=t,  S=1] &  (\text{A2})\\
        &= 0
    \end{align*}
    and so \[\theta(t) = \mathbb{E}_{\bX}\left[ \mu(\bX,t,0) - \mu(\bX,t,1) \right] = 0.\]
    Thus, if $\theta(t) \neq 0$ and A1 -- A3, A5 hold, then by the contrapositive of the above, A4 is violated; similarly, if $\theta(t) \neq 0$ and A1, A2, A4, A5 hold, then A3 is violated. \\
    
    \noindent \textbf{For the proof of (3),} we want to show that $\E[Y(t)|\bX=\bx]$ is identified under (A1-A3, A5) or (A1, A2, A4, A5). We proceed by considering two cases.\\
    \textbf{Case 1:} Assumptions A1-A3, A5 holds:
    \begin{align*}
    \E[Y(t)|\bX=\bx] &= \E[Y(t)|\bX=\bx,S=1] & (\text{A3}) \\
    &= \E[Y(t)|\bX=\bx,T=t,S=1] & (\text{A2})\\
    &= \E[Y |\bX=\bx,T=t,S=1] & (\text{A1})\\
    &= \mutexp.
    \end{align*}
    \\
    \textbf{Case 2:} Assumptions A1, A2, A4, A5 holds:
    \begin{align*}
        \E[Y(t)|\bX=\bx] &= \E[Y(t)\mid\bX=\bx,S=1]P(S=1\mid\bX=\bx) \\&\;\;\;\;+ \E[Y(t)\mid\bX=\bx,S=0]P(S=0\mid\bX=\bx)\\
        &= \E[Y(t) \mid\bX=\bx, T=t, S=1]P(S=1\mid\bX=\bx) \\&\;\;\;\;+ \E[Y(t)\mid\bX=\bx, T=t, S=0]P(S=0\mid\bX=\bx) & (\text{A2 \& A4})\\
         &= \E[Y \mid\bX=\bx, T=t, S=1]P(S=1\mid\bX=\bx) \\&\;\;\;\;+ \E[Y\mid\bX=\bx, T=t, S=0]P(S=0\mid\bX=\bx) & (\text{A1})\\
         &= \mutexp P(S=1\mid\bX=\bx) + \mutobs P(S=0\mid\bX=\bx). 
    \end{align*}
Hence, it is identified.
\hfill$\square$

\subsection{Proofs of Theorems \ref{thm:dml_theta} and \ref{thm:dml_pate}}
\subsubsection{Overview}
In this section, we prove Theorems \ref{thm:dml_theta} and \ref{thm:dml_pate}. In both cases, the approach will be to show that the estimators referenced therein can be derived from certain score functions to be defined later. By showing these scores to satisfy certain assumptions, we will be able to apply Theorems 3.1 and 3.2 and Corollary 3.1 from \cite{chernozhukov2018double} and show that the estimators are unbiasedness and asymptotic normality. More specifically, the score need to satisfy unbiasedness, linearity, differentiability, Neyman orthogonality, identifiability, boundedness of moments, and boundedness of approximation rates -- all to be defined below. Much of our language and proof technique is drawn from the proofs of Theorems 5.1 and 5.2 of \cite{chernozhukov2018double}.

The rest of this section proceeds as follow: first, we define requisite notations and conditions on the data generative model. We, then, state the two conditions\footnote{the \cite{chernozhukov2018double} paper refers to this as `assumptions', though we want to emphasize that we will verify -- not assume -- that the conditions therein hold.} (also proposed in \cite{chernozhukov2018double}) that must be satisfied in order to invoke the theory leading to our main results. We then prove a lemma that will be useful in our proofs and lastly proceed to the proofs of Theorems \ref{thm:dml_theta} and \ref{thm:dml_pate}, which consist of verifying the two assumptions above.
\subsubsection{Setup and Notation}
We write the data generative model as follows:
\begin{align}
    Y &= \alpha_0(T, S, \bX) + \epsy, & \EOm[\epsy \mid T, S, \bX] = 0 \label{eq:dgp_y} \\
    T &= \beta_0(S, \bX) + \mathcal{E}_T, & \EOm[\mathcal{E}_T \mid S, \bX] = 0 \label{eq:dgp_t} \\
    S &= \gamma_0(\bX) + \mathcal{E}_S, & \EOm[\mathcal{E}_S \mid \bX] = 0 \label{eq:dgp_s}
\end{align}

Further, recall that we defined $\Om = (Y, T, S, \bX)$. We now characterize $\Om$ as taking values in the measurable space $(Z, \mathcal{F}_{\Omega})$ according to a probability measure $P \in \mathcal{P}_N$ (while in the main text, we use $n$ to denote the total sample size, here we use $N$ for agreement with the notation of \cite{chernozhukov2018double}). Also, let $\pq{f} = \pq{f(\Omega)} = (\int |f(\omega)|^qd {P}(\omega))^{1/q}$ denote the $L^q(P)$ norm with respect to the distribution of the data. For a given a vector $\mathbf{f} = (f_1, \dots, f_k)$, we, further, define $\pq{\mathbf{f}} := \max_{1\leq \ell \leq k}\pq{f_\ell}$. \footnote{Other related technical details (e.g. measurability considerations) are as in \cite{chernozhukov2018double}.}

Note that the setup must individually apply to \emph{each} of the Theorems \ref{thm:dml_theta} and \ref{thm:dml_pate}. However, the constants and the like may differ from one proof to another.

Consider scores $\Psi(\Omega;\theta,\eta)$ that are functions of the parameter of our interest $\theta$ and the vector of nuisance parameter $\eta = (\alpha, \beta, \gamma)$. Here, each of $\alpha, \beta, \gamma$ are $P$-square integrable functions with co-domains $\mathbb{R}$, $(\eps_\beta, 1 - \eps_\beta)$, and $(\eps_\gamma, 1 - \eps_\gamma)$, respectively, where $\eps_\beta, \eps_\gamma \in (0, 1/2)$. For convenience, we further define $\eps:= \min(\eps_\beta, \eps_\gamma)$.

 For any integer $q>2$, we further characterize the nuisance set $\mathcal{W}_N$ as the set of $\eta = (\alpha, \beta, \gamma)$ satisfying: 
\begin{flalign}
    &\pq{\eta - \eta_0} \leq C \label{eq:nuisance_bound_q} \\
    &\ptwo{\eta - \eta_0} \leq \delta_N \label{eq:nuisance_bound_2} \\
    &\pq{\beta_0(S, \bX) - 1/2} \leq 1/2 - \eps_\beta,\quad \forall S \in \{0, 1\} \label{eq:overlap_beta} \\
    &\pq{\gamux - 1/2} \leq  1/2 - \eps_\gamma \label{eq:overlap_gamma} \\
    &\ptwo{\alpha(T) - \alpha_0(T)} \times \ptwo{\beta - \beta_0} \leq \delta_NN^{-1/2} ,\quad \forall T \in \{0, 1\} \label{eq:prog_prop_rate} \\
    &\ptwo{\alpha(T) - \alpha_0(T)} \times \ptwo{\gamma - \gamma_0} \leq \delta_NN^{-1/2} ,\quad \forall T \in \{0, 1\}. \label{eq:prog_selec_rate}
\end{flalign}
In the above, for notational convenience, we let $\ptwo{\alpha(T) - \alpha_0(T)}$ denote the maximum of $\ptwo{\alpha(T, 0, \bX) - \alpha_0(T, 0, \bX)}$ and $\ptwo{\alpha(T, 1, \bX) - \alpha_0(T, 1, \bX)}$. Similarly, $\ptwo{\beta - \beta_0}$ denotes the maximum of $\ptwo{\beta(0, \bX) - \beta_0(0, \bX)}$ and $\ptwo{\beta(1, \bX) - \beta_0(1, \bX)}$ and so for $\ptwo{\gamma - \gamma_0}$ as well.

$\eta_0$ denotes the true value of $\eta \in \mathcal{W}_N$. The true values of the nuisance parameters are, for $t, s \in \{0, 1\}$: $\alpha_0(t, s, \bX)$, $ \beta_0(s, \bX)$, and $\gamuxz$, where $\alpha_0(t, s, \bX) = \mathbb{E}[Y \mid T = t, S = s, \bX]$, $\beta_0(s, \bX) = \mathrm{P}(T = 1 \mid S = s, \bX)$, and $\gamuxz = \mathrm{P}(S = 1 \mid \bX)$, as defined in Equations \eqref{eq:dgp_y} -- \eqref{eq:dgp_s}. 

We now outline some additional regularity conditions. Start by taking $(\delta_N)_{N=1}^\infty$ and $(\Delta_N)_{N=1}^\infty$ to be sequences of non-negative constants monotonically decreasing towards 0; $c, C, \epsilon$, and $q$ to be strictly positive constants with $q > 2$; and $K \geq 2$ to be an integer that, for simplicity, evenly divides $N$. Lastly, we let $\{\mathcal{P}_N\}_{N \geq 1}$ be some sequence of sets of probability distributions $P$ of $\Om$ on $Z$. We then require the following regularity conditions to hold.

\textbf{Assumption S1. Regularity conditions.} For all probability laws $P \in \mathcal{P}$ for the tuple $(Y, T, S, \bX)$, the following conditions hold:
\begin{enumerate}
    \item Equations \eqref{eq:dgp_y} -- \eqref{eq:dgp_s} hold, with $T, S \in \{0, 1\}$
    \item $\pq{Y} \leq C$
    \item $\mathrm{P}(\eps_\beta \leq \beta(s, \bX) \leq 1 - \eps_\beta) = 1$ for $s \in \{0, 1\}$
    \item $\mathrm{P}(\eps_\gamma \leq \gamma(\bX) \leq 1 - \eps_\gamma) = 1$
    \item $\ptwo{\epsy} \geq c$
    \item $\|\EOm[\epsy ^ 2 \mid \bX]\|_{P, \infty} \leq C$
    \item Given a random subset $I$ of $\{1\dots N\}$ of size $N / K$, the nuisance parameter estimator $\hat{\eta}_0 = \hat{\eta}_0((\Omega_i)_{i \in I^c})$ belongs to the set $\mathcal{W}_N$ with $P$-probability no less than $1 - \Delta_N$, where $\eta_0 \in \mathcal{W}_N$.
\end{enumerate}

Here, we consider linear scores such that $\Psi(\Omega; \theta, \eta)$ can be decomposed as follows:
\begin{equation}
\Psi(\Omega; \theta, \eta) = \Psi^a(\Omega;\eta) \; \theta + \Psi^b(\Omega;\eta),\label{eq:psi_linear}
\end{equation}
where $\theta$ denotes the parameter of interest. Next, we define a pathwise G\^{a}teaux derivative map over the score functions:
\begin{equation}
    D_r[\eta - \eta_0] := \partial_r \mathbb{E}_{\Omega}[ \Psi(\Omega; \theta_0, \eta_0 + r(\eta - \eta_0) ], \;\; \forall \eta \in \mathcal{W}_N.
\end{equation} 
Similar to \cite{chernozhukov2018double}, we define: 
\begin{equation}
    \partial_\eta \mathbb{E}_{\Omega}[ \Psi(\Omega; \theta_0, \eta_0)][\eta - \eta_0 ] := D_0[\eta - \eta_0],
\end{equation} for convenience.

Now, we summarize conditions sufficient for verifying double machine learning properties and applying Theorem 3.1, 3.2 and Corollary 3.1 from \cite{chernozhukov2018double}.

\subsubsection{Conditions to Verify for Double Machine Learning Properties}

\paragraph{\textbf{Condition 1 (Linear scores with approximate Neyman Orthogonality)}}
For all $N \geq 3$ and $P \in \mathcal{P}_N$, the following conditions hold: 
\begin{enumerate}
\item \textit{Score Unbiasedness: }The true parameter value $\theta_0$ satisfies $\mathbb{E}_{\Omega}[\Psi(\Om; \eta_0, \theta_0)] = 0$. 
\item \textit{Score Linearity: }The score $\Psi$ is linear in the sense of Equation \eqref{eq:psi_linear}. 
\item \textit{Score Differentiability: }The map $\eta \rightarrow \mathbb{E}_{\Omega} [\Psi(\Om; \theta, \eta)]$ is twice continuously G\^{a}teaux-differentiable. 
\item \textit{Neyman Orthogonality: }The score obeys the Neyman orthogonality condition or, more generally, the Neyman $\lambda_N$ near-orthogonality condition at $(\theta_0, \eta_0)$ with respect to the nuisance realization set $\mathcal{W}_N \subset W$ for

\[\lambda_N := \sup_{\eta \in \mathcal{W}_N}\|\partial_\eta \mathbb{E}_{\Omega}  \Psi(\Om; \theta_0, \eta_0)[\eta - \eta_0]\| \leq \delta_N N ^ {-1/2}.\]

\item \textit{Score Identifiability: }The identification condition holds; namely, the singular values of the matrix
\[J_0 := \EOm[\Psi^a(\Om; \eta_0)]\]
are between $c_0$ and $c_1$. 
\end{enumerate}

\paragraph{\textbf{Condition 2 (Score regularity and quality of nuisance parameter estimators)}}
For all $N \geq 3$ and $P \in \mathcal{P}_N$, the following conditions hold:
\begin{enumerate}
\item \textit{Sufficient Mass of Realization Set: }Given a random subset $I$ of $\{1\dots N\}$ of size $n = N / K$, the nuisance parameter estimator $\hat{\eta}_0 = \hat{\eta}_0((\Om_i)_{i \in I^c})$ belongs to the realization set $\mathcal{W}_N$ with probability at least $1 - \Delta_N$, where $\mathcal{W}_N$ contains $\eta_0$. 
\item \textit{Boundedness of Score Moments: }The moment conditions hold: 
\begin{gather*}
    m_N := \sup_{\eta \in \mathcal{W}_N}(\EOm\|\Psi(\Om; \theta_0, \eta)\|^q)^{1/q} \leq c_1; \\
    m'_N := \sup_{\eta \in \mathcal{W}_N}(\EOm\|\Psi^a(\Om; \eta)\|^q)^{1/q} \leq c_1; 
\end{gather*}

\item \textit{Boundedness of Score Approximation Rates: }The following conditions on the statistical rates $r_N, r'_N$, and $\lambda'_N$ hold:
\begin{gather*}
    r_N := \sup_{\eta \in \mathcal{W}_N} \|\EOm[\Psi^a(\Om; \eta)] - \EOm[\Psi^a(\Om; \eta_0)]\| \leq \delta_N, \\ 
    r'_N := \sup_{\eta \in \mathcal{W}_N} (\EOm[\|\Psi(\Om; \theta_0, \eta) - \Psi(\Om; \theta_0, \eta_0)\| ^ 2]) ^ {1/2} \leq \delta_N, \\
    \lambda'_N := \sup_{r \in (0, 1), \eta \in \mathcal{W}_N} \|\partial_r ^ 2 \EOm[\Psi(\Om; \theta_0, \eta_0 + r(\eta - \eta_0))]\| \leq \delta_N / \sqrt{N}.
\end{gather*}
\item \textit{Non-Degeneracy of Score Variance: }The variance of the score $\Psi$ is non-degenerate: all eigenvalues of the matrix 
\[\EOm[\Psi(\Om; \theta_0, \eta_0)\Psi(\Om; \theta_0, \eta_0)']\]
are bounded from below by $c_0$.
\end{enumerate}

For each of Theorems \ref{thm:dml_theta} and \ref{thm:dml_pate} we will replace the sequence $(\delta_N)_{N=1}^\infty$ by a sequence $(\delta'_N)_{N=1}^\infty$ defined by $\delta'_N := K(\max(\delta_N , N^{-(\min((1-2/q), 1)/2)}))$ where $K$ is a constant unique to each theorem. Note that, as $\delta'_N$ is still lower bounded by $N^{-1/2}$; because it is also lower-bounded by $N^{-(\min((1-2/q), 1)/2)}$, we are able to apply Theorem 3.2 from \cite{chernozhukov2018double}, which allows us to construct a consistent variance estimator. 

\subsubsection{Norm Bounds for Nuisance Parameters}
Before proceeding to the proofs of Theorems \ref{thm:dml_theta} and \ref{thm:dml_pate}, it will be useful to establish the following lemma. 
\begin{lemma}(Norm Bounds for Nuisance Parameters)\label{lemma:norm_bounds}
    Given Assumption S1, we have that the following bounds hold:
    \begin{gather*}
        \pq{\alpha_0(t, s, \bX)} \leq C/\eps^{2/q},\quad \forall t,s \in \{0, 1\} \\
        \pq{\alpha_0(t, s, \bX) - \alpha(t, s, \bX)} \leq C/\eps^{2/q},\quad \forall t,s \in \{0, 1\} \\
        \ptwo{\alpha_0(t, s, \bX) - \alpha(t, s, \bX)} \leq \delta_N/\eps,\quad \forall t,s \in \{0, 1\}
    \end{gather*}
\end{lemma}
\begin{proof}
\begin{align*}
\color{black}{\pq{\alpha_0(T, S, \bX)}} &= \color{black}{(\EOm|\alpha_0(T, S, \bX)| ^ q)^{1/q}} \\
&\color{black}{= \left(\int_{\bx} \sum_{t, s \in \{0, 1\}} |\alpha_0(t, s, \bx)| ^ q ~ P(T = t, S = s \mid \bX=\bx) p(\bx) d\bx \right)^{1/q}} \\
&\color{black}{\geq \left(\int_{\bx} \sum_{t, s \in \{0, 1\}} |\alpha_0(t, s, \bx)| ^ q ~ \eps^{2/q} p(\bx) d\bx \right)^{1/q}} \\
&\color{black}{= \eps^{2/q} \left(\int_{\bx} \sum_{t, s \in \{0, 1\}} |\alpha_0(t, s, \bx)| ^ q ~  p(\bx) d\bx \right)^{1/q}} \\
&\color{black}{\geq \eps^{2/q} \left(\int_{\bx} \max_{t, s \in \{0, 1\}} |\alpha_0(t, s, \bx)| ^ q ~  p(\bx) d\bx \right)^{1/q}} \\
&= \color{black}{\eps^{2/q} \left( \mathbb{E}_{\bX}\left|\max_{t, s \in \{0, 1\}} \alpha_0(t, s, \bX)\right| ^ q \right)^{1/q}} \\
&\color{black}{\overset{(\textrm{By Jensen's Inq.})}{\geq} \eps^{2/q} \max_{t, s \in \{0, 1\}} \left( \mathbb{E}_{\bX}|\alpha_0(t, s, \bX)| ^ q \right)^{1/q}} \\
&\color{black}{= \eps^{2/q} \max_{t, s \in \{0, 1\}} \pq{\alpha_0(t, s, \bX)}}
\end{align*}
where the second inequality comes from the bounds on the ranges of $\beta_0$ and $\gamma_0$.\\  
\textcolor{black}{Thus, because $\pq{\alpha_0(T, S, \bX)}\leq \pq{Y}$} \textcolor{black}{$\leq C$} by Jensen's and Assumption S1(b), we have that 
\begin{equation}
\label{eq:g0bound}
\textcolor{black}{\pq{\alpha_0(t, s, \bX)} \leq C / \eps^{2 / q}\qquad \forall t,\, s \in \{0, 1\}}
\end{equation}

Similarly, for any $\eta \in \mathcal{W}_N$:

\begin{align*}
    \color{black}{C} &\color{black}{\geq\pq{\alpha(T, S, \bX) - \alpha_0(T, S, \bX)}} \\
      &\color{black}{\geq \left(\sum_{t, s \in \{0, 1\}} \EOm [|\alpha(t, s, \bX) - \alpha_0(t, s, \bX)| ^ q] ~ \POm(T = t, S = s \mid \bX)\right)^{1/q}} \\
      &\geq \color{black}{\eps^{2/q} \left(\sum_{t, s \in \{0, 1\}} \EOm |\alpha(t, s, \bX) - \alpha_0(t, s, \bX)| ^ q\right)^{1/q}} \\
      &\color{black}{\geq \eps^{2/q} \left(\max_{t, s \in \{0, 1\}} \EOm |\alpha(t, s, \bX) - \alpha_0(t, s, \bX)| ^ q\right)^{1/q}} \\
      &\color{black}{\geq \eps^{2/q} \max_{t, s \in \{0, 1\}} \pq{\alpha(t, s, \bX) - \alpha_0(t, s, \bX)} }
\end{align*}
where the first inequality follows from the assumption in Inequality \eqref{eq:nuisance_bound_q}, and so
\begin{equation}
\label{eq:alpha_diff_bound_p}
\color{black}{\pq{\alpha(t, s, \bX) - \alpha_0(t, s, \bX)} \leq C / \eps^{2 / q}\qquad \forall t,\, s \in \{0, 1\}}
\end{equation}
Repeating the last set of calculations with respect to the $L_2(P)$ norm and then applying Inequality \eqref{eq:nuisance_bound_2} yields the analogous statement:
\begin{equation}
\label{eq:alpha_diff_bound_2}
\color{black}{\ptwo{\alpha(t, s, \bX) - \alpha_0(t, s, \bX)} \leq \delta_N / \eps \qquad \forall t,\, s \in \{0, 1\}}
\end{equation}
\end{proof}
\subsubsection{Proof of Theorem \ref{thm:dml_theta}}
\setcounter{theorem}{3}
\begin{theorem}
Let Assumptions (A1-A6) be satisfied. Then:
\begin{align*}
    \sqrt{n}\thetahat(t) \overset{d}{\rightarrow} \mathcal{N}(0, \sigmasq(t)),
\end{align*}
where: $\sigmasq(t) = \E\left[(\Psiithat - \thetahat(t))^2\right]$ and the expectation is taken over $\Om$. \\In addition, the naive estimator for the variance is consistent with the first-stage estimates used in place of the true nuisance parameters:
$$\sigmasqhat(t) = \frac{1}{n}\sum_{i=1}^N(\Psiithat - \thetahat(t))^2 \overset{p}{\rightarrow} \sigmasq(t),$$
Finally, it follows that an uniformly valid asymptotic $1-\alpha$ confidence interval is: $$\left[\thetahat(t) \pm \Phi^{-1}(1-\alpha/2)\sqrt{\sigmasqhat(t)/n}\right].$$
\end{theorem}
\paragraph{\textit{Proof:}} Recall, in Section~\ref{sec:testing}, $\theta(t) = \mathbb{E}[\mutobs - \mutexp]$ and $\thetahat(t) = \frac{1}{N}\sum_{i=1}^N \Psiithat,$
where:
\begin{equation*}
\Psiit = \muitobs - \muitexp + 
\frac{\Tit (1-S_i)}{\eitobs(1-\pri)}(Y_i - \muitobs) - \frac{\Tit(S_i)}{\eitexp\pri}(Y_i - \muitexp), \text{and}
\end{equation*}
\begin{equation*}
\Psiithat = \muitobshat - \muitexphat + 
\frac{\Tit (1-S_i)}{\eitobshat(1-\prihat)}(Y_i - \muitobshat) - \frac{\Tit(S_i)}{\eitexphat\prihat}(Y_i - \muitexphat).
\end{equation*}
For this proof, we will focus on the case $t = 1$, however, this proof generalizes trivially for $t=0$. In this case, the parameter of interest $\theta_0$ is $\theta(1)$. 

Now, consider the score function
\begin{align*}\Psi(\Om; \theta, \eta) = & \; \phi(1) - \theta(1)\\
= &\underbrace{\frac{T(1 - S)}{\beta(0, \bX)(1 - \gamux)}(Y - \alpha(1, 0, \bX))}_{(a)} - \\ 
&\underbrace{\frac{TS}{\beta(1, \bX)\gamux}(Y - \alpha(1, 1, \bX))}_{(b)} + \\ 
&\underbrace{\alpha(1, 0, \bX) - \alpha(1, 1, \bX) -\theta}_{(c)} \end{align*}
and note that it is linear in the sense of Equation \eqref{eq:psi_linear} with $\Psi^a = -1$, with the true value of the parameter of interest $\theta$ being $\color{black}{\theta_0}$. Based of off \cite{chernozhukov2018double}'s Theorem 3.1 and 3.2 (on pages C26-27), and corollaries 3.1 and 3.2 (on page C27), if conditions 1 and 2 are satisfied then 
$\sqrt{n}\thetahat(t) \overset{d}{\rightarrow} \mathcal{N}(0, \sigmasq(t))$ and
$\sigmasqhat(t) = \frac{1}{n}\sum_{i=1}^N(\Psiithat - \thetahat(t))^2 \overset{p}{\rightarrow} \sigmasq(t)$.

We will now proceed by verifying Conditions 1 and 2.

\subsection*{Verification of Condition 1 for Theorem~\ref{thm:dml_theta}}
\textbf{Condition 1.1.} We begin by verifying the moment condition 
\[\EOm[\Psi(\Om; \theta_0, \eta_0)] = 0.\]

We will do so by showing that the expectation of each of $(a)$, $(b)$, and $(c)$ -- all with true values plugged in -- are each equal to 0.

We start by focusing on $(a)$ and plug-in true values $\eta_0$. As a consequence of data generative mechanism, $\alpha_0(t,s,\bX) = \mathbb{E}[Y | T=t, S=s, \bX]$, $\beta_0(s,\bX) = P(T=1 | S=s, \bX)$ and $\gamma_0(\bX) = P(S=1 | \bX)$. Thus the first term can be rewritten:

\begin{eqnarray*}
    \color{black}&{\mathbb{E}_{YTSX}\left[\frac{T(1 - S)}{\mathrm{P}(T = 1 \mid S = 0, \bX)\mathrm{P}(S = 0 \mid \bX)}(Y - \mathbb{E}[Y | T=1, S=0, \bX]) \right]} &= \\ 
    &\color{black}{\mathbb{E}_{X}\mathbb{E}_{YTS|X}\bigg[\frac{T(1 - S)(Y - \mathbb{E}[Y | T=1, S=0, \bX])}{\mathrm{P}(T = 1 \mid S = 0, \bX)\mathrm{P}(S = 0 \mid \bX)}\mid \bX \bigg]} &= \\
    &\color{black}{\mathbb{E}_{X}\bigg[\frac{(\mathbb{E}(Y\mid T = 1, S = 0, \bX)- \mathbb{E}[Y | T=1, S=0, \bX]))\mathrm{P}(T = 1 \mid S = 0, \bX)\mathrm{P}(S = 0 \mid \bX)}{\mathrm{P}(T = 1 \mid S = 0, \bX)\mathrm{P}(S = 0 \mid \bX)}\bigg]}& = 0
\end{eqnarray*}
Analogously, $\color{black}{\mathbb{E}_{YTSX}[(b)]_{\eta = \eta_0} = 0}$ as well.
Now, we show that $\mathbb{E}_{YTSX}[(c)]_{\eta = \eta_0} = 0$. Evaluating $\mathbb{E}[(c)]$ at $\alpha_0$ and $\theta_0$ we have 
\begin{align*}
    \color{black}{\mathbb{E}_{YTSX}[(c)]_{\eta = \eta_0} }=
    &\color{black}{\E_{YTSX}[\E[Y \mid T = 1, S = 0, \bX] - \E[Y \mid T = 1, S = 1, \bX] +} \\ &\color{black}{\E_X\E[Y \mid T = 1, S = 1, \bX] - \E_X\E[Y \mid T = 1, S = 0, \bX]]} \\
    = &\color{black}{\E_{\bX}[\E[Y \mid T = 1, S = 0, \bX] - \E[Y \mid T = 1, S = 1, \bX]] +} \\ &~\color{black}{\E_{\bX}\E[Y \mid T = 1, S = 1, \bX] - \E_{\bX}\E[Y \mid T = 1, S = 0, \bX]}\\
    = &\color{black}{0}
\end{align*}
and thus the moment condition holds.
\paragraph{\textbf{Condition 1.5.}} We now verify the score is Neyman-orthogonal i.e. $\sup_{\eta \in \mathcal{W}_N} \partial_\eta \mathbb{E}_{\Omega}[ \Psi(\Omega; \theta_0, \eta_0)][\eta - \eta_0 ]\leq \delta_N / N^{1/2}$. 
Recall, that \begin{equation}
    \partial_\eta \mathbb{E}_{\Omega}[ \Psi(\Omega; \theta_0, \eta_0)][\eta - \eta_0 ] := D_0[\eta - \eta_0],
\end{equation} where $D_0[\eta - \eta_0]$ is a G\^{a}teaux derivative at $r=0$ defined as 
\begin{equation}
    D_r[\eta - \eta_0] := \partial_r \mathbb{E}_{\Omega}[ \Psi(\Omega; \theta_0, \eta_0 + r(\eta - \eta_0) ], \;\; \forall \eta \in \mathcal{W}_N.
\end{equation} 

For any $\eta \in \mathcal{W}_N$, the G\^{a}teaux derivative in the direction $\eta - \eta_0$ is:
\begin{align*}
\partial_\eta \EOm [\Psi(\Om; \theta_0, \eta_0)][\eta - \eta_0] = &\EOm[\alpha(1, 0, \bX) -\alpha_0(1, 0, \bX)\big] - \\
&\EOm[\alpha(1, 1, \bX) - \alpha_0(1, 1, \bX)] + \\
&\EOm \big[\frac{(1-S) T (\alpha_0(1, 0, \bX)-\alpha(1, 0, \bX))}{\beta_0(0, \bX) (1-\gamma_0(\bX))}\big] - \\
&\EOm \big[\frac{S T (\alpha_0(1, 1, \bX) - 
\alpha(1, 1, \bX))}{\beta_0(1, \bX) \gamma_0(\bX)}\big] - \\ 
&\EOm\big[\frac{(1-S) T (Y-\alpha_0(1, 0, \bX)) (\beta(0, \bX) - \beta_0(0, \bX))}{\beta_0(0, \bX)^2 (1-\gamma_0(\bX))}\big] - \\
&\EOm\big[\frac{(1-S) T (Y-\alpha_0(1, 0, \bX)) (\gamma_0(\bX)-\gamma(\bX))}{\beta_0(0, \bX) (1-\gamma_0(\bX))^2}\big] + \\
&\EOm\big[\frac{S T (\beta(1, \bX)-\beta_0(1, \bX)) (Y-\alpha_0(1, 1, \bX))}{\beta_0(1, \bX)^2 \gamma_0(\bX)}\big] + \\
&\EOm\big[\frac{S T (Y-\alpha_0(1, 1, \bX)) (\gamma(\bX)-\gamma_0(\bX))}{\beta_0(1, \bX) \gamma_0(\bX)^2}\big].
\end{align*}

\textcolor{black}{By iterated expectation, the first four terms cancel (the first with the third, and the second with the fourth). The remaining four terms each equal 0 by iterated expectation as $ST(Y - \alpha_0(1, 1, \bX))$ and $(1 - S)T(Y - \alpha_0(1, 0, \bX))$ can be shown to each equal 0 in the same way as the moment condition was verified. Thus, the score is Neyman-orthogonal.}

\textbf{Conditions 1.2, 1.3, 1.4}
\textcolor{black}{We have already shown the score to be linear and the map $\eta \rightarrow \EOm [\Psi]$ can trivially be shown to be twice G\^{a}teaux differentiable, satisfying Conditions 1.2. and 1.3. Lastly, because $\Psi^a = -1$, the identification requirement of Conditions 3.1 (4) -- that the singular values of $\EOm[\Psi^a]$ (here, 1) are bounded -- is satisfied as well.}

Thus, all of Condition 1 is satisfied. 

\subsection*{Verification of Condition 2 for Theorem~\ref{thm:dml_theta}}
\paragraph{\textbf{Condition 2.1.}} Condition 2.1. is automatically satisfied by Assumption S1(g) and the construction of $\mathcal{W}_N$. 

\paragraph{\textbf{Condition 2.2.}}
Rewrite the score:
\begin{align*}
    \Psi(\Om; \theta, \eta) = 
    &\alpha(1, 0, \bX)\left(1 - \frac{T(1 - S)}{\beta(0, \bX)(1 - \gamma(\bX))}\right) - \\
    & \alpha(1, 1, \bX)\left(1 - \frac{TS}{\beta(1, \bX)\gamma(\bX)}\right) + \\
    &Y\left(\frac{T(1 - S)}{\beta(0, \bX)(1 - \gamma(\bX))} - \frac{TS}{\beta(1, \bX)\gamma(\bX)}\right)- \theta
\end{align*}

We can thus bound $\pq{\Psi(\Om; \theta_0, \eta)}$:
\begin{equation}\label{eq:psi_theta_norm}
    \begin{split}
    \color{black}{\pq{\Psi(\Om; \theta_0, \eta)}} \leq & \color{black}{\pq{\alpha(1, 0, \bX)\left(1 - \frac{T(1 - S)}{\beta(0, \bX)(1 - \gamma(\bX))}\right)}} + \\
    & \color{black}{\pq{\alpha(1, 1, \bX)\left(1 - \frac{TS}{\beta(1, \bX)\gamma(\bX)}\right)}} + \\
    & \color{black}{\pq{Y\left(\frac{T(1 - S)}{\beta(0, \bX)(1 - \gamma(\bX))} - \frac{TS}{\beta(1, \bX)\gamma(\bX)}\right)} + |\theta_0|}  \\
    \leq &\color{black}{(1 + \eps^{-2})\left(\pq{\alpha(1, 0, \bX)} + \pq{\alpha(1, 1, \bX)}\right)} +
    2\pq{Y}/\eps ^ 2 + |\theta_0| \\
    \leq & \color{black}{(1 + \eps^{-2})\left(\pq{\alpha(1, 0, \bX) - \alpha_0(1, 0, \bX)} + \pq{\alpha(1, 1, \bX) - \alpha_0(1, 1, \bX)}\right) +} \\
    & \color{black}{(1 + \eps^{-2})\left(\pq{\alpha_0(1, 0, \bX)} + \pq{\alpha_0(1, 1, \bX)}\right) +
    2\pq{Y}/\eps ^ 2 + |\theta_0|}
    \end{split}
\end{equation}
where we use the fact that $T$ and $S$ are binary and that $\beta(0, \bX)$, $\beta(1, \bX)$, $\gamma(\bX)$, and $1 - \gamma(\bX)$ are all bounded below by $\eps$.

We now bound each of the terms of \eqref{eq:psi_theta_norm} in turn. The first two terms are bounded by Lemma \ref{lemma:norm_bounds}, by Theorem \ref{thm:id} we have that $\theta_0 = 0$ under the null, and by Assumption S1(b) we have that $\color{black}{\pq{Y} \leq C}$. Thus, we can extend the bound in \eqref{eq:psi_theta_norm} to conclude:

\begin{align*}
    \color{black}{\pq{\Psi(\Om; \theta_0, \eta)}} &\leq \color{black}{4C(1 + \eps^{-2})/\eps^{2/q} + 2C/\eps ^ 2}
\end{align*}
which gives the required bound on $m_n$ in Condition 2.2. \textcolor{black}{Because $\Psi^a = -1$, we also immediately have that $m'_N = 1$ and so all of Condition 2.2. is satisfied.}

\paragraph{\textbf{Condition 2.3}}
As $\Psi^a$ does not involve any nuisance parameters, \[\color{black}{\|\EOm[\Psi^a(\Om; \eta) - \Psi^a(\Om; \eta_0)]\| = |-1 + 1| = 0 \leq \delta_N}\]
and so the bound on the rate $r_N$ is satisfied. 

To bound the rate $r'_N$, we first bound:
\color{black}
\begin{align*}
\ptwo{\Psi\left(\Om; \theta_{0}, \eta\right)-\Psi\left(\Om, \theta_{0}, \eta_{0}\right)} &\leq \underbrace{\ptwo{\alpha(1,0,\bX)-\alpha_{0}(1,0,\bX)}}_{\mathcal{I}_1} + \\
&\quad\underbrace{\ptwo{\alpha(1,1,\bX)-\alpha_{0}(1, 1,\bX)}}_{\mathcal{I}_2} + \\
&\quad\underbrace{\ptwo{\frac{T S\left(Y-\alpha\left(1,1,\bX\right)\right)}{\beta\left(1, \bX\right) (1 - \gamux)}-\frac{T S\left(Y-\alpha_{0}\left(1, 1, \bX\right)\right)}{\beta_{0}\left(1,\bX\right) \gamma_{0}(\bX)}}}_{\mathcal{I}_3} + \\
&\quad\underbrace{\ptwo{\frac{T(1 - S)(Y-\alpha(1,0, \bX))}{\beta\left(0, \bX\right)(1-\gamma(\bX))}-\frac{T(1 - S)\left(Y-\alpha_{0}(1, 0, \bX)\right)}{\beta_0(0, \bX)\left(1-\gamma_{0}(\bX)\right)}}}_{\mathcal{I}_4}
\end{align*}

By Lemma \ref{lemma:norm_bounds}, we can bound each of the first two terms $\mathcal{I}_1$ and $\mathcal{I}_2$ by $\delta_N/\eps$. \color{black} The fourth term, $\mathcal{I}_4$, can then be bounded:
\begin{align*}
\begin{split}
\mathcal{I}_4 \leq{} & \color{black}{\eps^{-4} \| T(1-S) \beta_0(0, \bX)\left(1 - \gamuxz)\right)(Y - \alpha(1, 0, \bX))-} 
\\ &\qquad \color{black}{T(1 - S) \beta(0, \bX)\left(1 - \gamux\right)(Y - \alpha_0(1, 0, \bX))\|_{P, 2}} \\
\leq &\color{black}{\eps^{-4} \|\beta_0(0, \bX)\left(1 - \gamuxz)\right)\left(\alpha_0(1, 0, \bX) + \epsy - \alpha(1, 0, \bX)\right) -} \\
 & \qquad \color{black}{\beta(0, \bX)(1 - \gamma(\bX)) \epsy\|_{P,2}} \\
\leq &\color{black}{\eps^{-4}\ptwo{\beta_0 \left(0, \bX\right)\left(1 -\gamuxz\right)\left(\alpha_{0}(1, 0, \bX) - \alpha(1, 0, \bX)\right)} +} \\
&\color{black}{\eps^ {-4} \ptwo{\epsy \left(\beta_0(0, \bX) \left(1 - \gamuxz)-\beta(0, \bX)(1 - \gamma(\bX)\right)\right)}} \\
\leq & \color{black}{\eps^{-4}\ptwo{\alpha_0(1, 0, \bX) - \alpha(1, 0, \bX)} +} \\
&\color{black}{\eps^{-4} \sqrt{C} \ptwo{\left(\beta_0(0, \bX) \left(1 - \gamuxz)-\beta(0, \bX)(1 - \gamma(\bX)\right)\right)}} \\
\leq & \color{black}{\eps^{-4}\delta_N\left(\eps^{-1} + \sqrt{C}C_{\delta}\right)}.
\end{split}
\end{align*}
In the first inequality we use the boundedness of the nuisance parameters $\beta(0, \bX)$, $\beta(1, \bX)$, $\gamux)$ and their true values. In the second, we use that $S$ and $T$ are binary, and that for $S = 0$ and $T = 1$, $T(1 - S)Y = \alpha_0(1, 0, \bX) + \epsy$. The third follows from the triangle inequality and the fourth from Assumption S1. The final inequality follows from Lemma \ref{lemma:norm_bounds} and the boundedness of $|\beta(0, \bX) - \beta_0(0, \bX)|$ and $|\gamux - \gamuxz|$ and involves the constant $C_\delta$ that is only a function of the first element of the sequence $(\delta_N)_{N=1}^{\infty}$. 

The third term, $\mathcal{I}_3$ can be bounded in analogous fashion, yielding the overall bound:
\begin{align*}\color{black}{2(\eps^{-1} + \eps^{-5} + \eps^{-4}\sqrt{C}C_{\delta})\delta_N} \leq {\delta'_N}\end{align*}
given that $K$ in the definition of $\delta'_N$ satisfies $K \geq 2(\eps^{-1} + \eps^{-5} + \eps^{-4}\sqrt{C}C_{\delta})$.

Lastly, to bound the rate $\lambda'_N$, we need to bound $|\partial^2 f(r)|$, where $f(r) := \EOm[\Psi(\Om; \theta_0, \eta_0 + r(\eta - \eta_0))]$ for $r \in (0, 1)$. For ease of legibility, we will write: $\gamma := \gamma(\bX)$, $\beta^s := \beta(s, \bX)$, and $\alpha^s := \alpha(1, s, \bX)$ for $s = 0, 1$ -- and similarly for the true values of the nuisance parameters.\vfill
For any $r \in (0, 1)$:
\begin{align*}
\partial^2 f(r) &= \EOm\bigg[\frac{2 (1-S) T (\beta^0-\beta^0_0)^2 (-r (\alpha^0-\alpha^0_0)-\alpha^0_0+Y)}{(r (\beta^0-\beta^0_0)+\beta^0_0)^3 (-r (\gamma-\gamma_0)-\gamma_0+1)}\bigg] \\
&+ \EOm\bigg[\frac{2 (1-S) T (\beta^0-\beta^0_0) (\gamma_0-\gamma) (-r (\alpha^0-\alpha^0_0)-\alpha^0_0+Y)}{(r (\beta^0-\beta^0_0)+\beta^0_0)^2 (-r (\gamma-\gamma_0)-\gamma_0+1)^2}\bigg] \\
&+ \EOm\bigg[\frac{2 (1-S) T (\gamma_0-\gamma)^2 (-r (\alpha^0-\alpha^0_0)-\alpha^0_0+Y)}{(r (\beta^0-\beta^0_0)+\beta^0_0) (-r (\gamma-\gamma_0)-\gamma_0+1)^3}\bigg] \\
&- \EOm\bigg[\frac{2 (1-S) T (\alpha^0_0-\alpha^0) (\beta^0-\beta^0_0)}{(r (\beta^0-\beta^0_0)+\beta^0_0)^2 (-r (\gamma-\gamma_0)-\gamma_0+1)}\bigg] \\
&- \EOm\bigg[\frac{2 (1-S) T (\alpha^0_0-\alpha^0) (\gamma_0-\gamma)}{(r (\beta^0-\beta^0_0)+\beta^0_0) (-r (\gamma-\gamma_0)-\gamma_0+1)^2}\bigg] \\
&- \EOm\bigg[\frac{2 S T (\beta^1-\beta^1_0)^2 (-r (\alpha^1-\alpha^1_0)-\alpha^1_0+Y)}{(r (\beta^1-\beta^1_0)+\beta^1_0)^3 (r (\gamma-\gamma_0)+\gamma_0)}\bigg] \\
&- \EOm\bigg[\frac{2 S T (\beta^1-\beta^1_0) (\gamma-\gamma_0) (-r (\alpha^1-\alpha^1_0)-\alpha^1_0+Y)}{(r (\beta^1-\beta^1_0)+\beta^1_0)^2 (r (\gamma-\gamma_0)+\gamma_0)^2}\bigg] \\
&- \EOm\bigg[\frac{2 S T (\gamma-\gamma_0)^2 (-r (\alpha^1-\alpha^1_0)-\alpha^1_0+Y)}{(r (\beta^1-\beta^1_0)+\beta^1_0) (r (\gamma-\gamma_0)+\gamma_0)^3}\bigg] \\
&+\EOm\bigg[\frac{2 S T (\beta^1-\beta^1_0) (\alpha^1_0-\alpha^1)}{(r (\beta^1-\beta^1_0)+\beta^1_0)^2 (r (\gamma-\gamma_0)+\gamma_0)}\bigg] \\
&+ \EOm\bigg[\frac{2 S T (\alpha^1_0-\alpha^1) (\gamma-\gamma_0)}{(r (\beta^1-\beta^1_0)+\beta^1_0) (r (\gamma-\gamma_0)+\gamma_0)^2}\bigg]
\end{align*}

\textcolor{black}{To upper bound $|\partial^2 f(r)|$, note that the expectation and the absolute value can be exchanged by Jensen's inequality, and the absolute value then distributed across the sum. Then, note that the absolute value of each denominator has an infimum over $(r, \bX)$ that is bounded away from 0\footnote{Each denominator is a polynomial whose absolute value is bounded away from 0 because each contains at least one term that is at most a function of $\gamma,\gamma_0,\beta^1,\beta^1_0,\beta^0,\beta^0_0$, each of which are bounded away from 0.} and so we can bound one over the absolute value of the denominator of each term all by one over the minimum of the infima of the absolute values of the denominators, which will be a constant depending solely on $\epsilon$.} In the below, we thus use $C'_{\epsilon}$ to denote a constant (possibly changing from line to line) that depends at most on $C$ and $\epsilon$. 

The absolute value of the ninth term can thus be bounded by:
\begin{align*}
    &\color{black}{C'_{\eps}\EOm\big[2 S T \left\lvert(\alpha_0 ^ 1- \alpha ^ 1) 
              (\beta^1-\beta_0^1)\right\rvert\big] \leq 
    C'_{\eps}\EOm\big[\left\lvert(\alpha_0 ^ 1- \alpha ^ 1) 
              (\beta^1-\beta_0^1)\right\rvert\big]} \\ &\leq 
    \color{black}{C'_{\eps}\left(\EOm(\alpha_0 ^ 1- \alpha ^ 1)^2\EOm(\beta^1-\beta_0^1) ^ 2\right)^{1/2} = 
    C'_{\eps} \times \|\alpha_0 ^ 1- \alpha ^ 1\|_{P, 2} \times \|\beta^1-\beta_0^1\|_{P, 2}}
\end{align*}
where we use the fact that $S$ and $T$ are binary in the first inequality, and Cauchy-Schwarz in the second. In analogous fashion, the absolute value of the fourth term can also be bounded by 
\[\color{black}{C'_{\eps} \times \|\alpha_0 ^ 1- \alpha ^ 1\|_{P, 2} \times \|\beta^1-\beta_0^1\|_{P, 2}}\]
and the absolute values of the fifth and tenth terms can be respectively bounded by:
\begin{gather*}
  C'_{\eps} \times \|\alpha_0 ^ 0- \alpha ^ 0\|_{P, 2} \times \|\gamma-\gamma_0\|_{P, 2} \\
  C'_{\eps} \times \|\alpha_0 ^ 1- \alpha ^ 1\|_{P, 2} \times \|\gamma-\gamma_0\|_{P, 2}
\end{gather*}

The absolute value of the first term can be bounded by:
\begin{align*}
    &~~ C'_{\eps}\EOm \big[2(1 - S)T|(\beta^0 - \beta^0_0)^2(-r(\alpha^0 - \alpha^0_0) - \alpha_0^0 + Y)|\big] \\
    &\leq  C'_{\eps}\EOm \big[2(1 - S)T|(\beta^0 - \beta^0_0)^2(-r(\alpha^0 - \alpha^0_0) + \epsy)|\big] \\
    &\leq  C'_{\eps}\left(\EOm \big[(1 - S)T(\beta^0 - \beta^0_0)^2|\epsy|\big] + \EOm\big[(1 - S)T(\beta^0 - \beta^0_0)^2|r(\alpha^0 - \alpha^0_0)|\big]\right) \\
    &\leq C'_{\eps}\left(\EOm \big[(1 - S)T|\epsy|\big] + \EOm\big[(1 - S)T|\beta^0 - \beta^0_0| \times |r(\alpha^0 - \alpha^0_0)|\big]\right) \\
    &\leq C'_{\eps}\EOm\big[|(\beta^0 - \beta^0_0)(\alpha^0 - \alpha^0_0)|\big] \\
    &\leq  C'_{\eps} \times \ptwo{\beta^0 - \beta^0_0} \times \ptwo{\alpha^0 - \alpha^0_0}
\end{align*}
where in the first inequality we use that fact that $(1 - S)T(Y - \alpha_0^0) = \epsy$, in the third inequality, we use the fact that $|\beta_0^0 - \beta^0| \leq 2$, and in the fourth inequality we use the facts that $S$ and $T$ are binary and $r$ is a constant in $(0, 1)$, and Assumption S1(f). The final inequality follows by Cauchy-Schwarz. 

Using similar arguments, the second term can eventually be bounded by 
\[C'_{\eps}\EOm\big[|(\beta^0 - \beta^0_0)(\alpha^0 - \alpha^0_0)|\big] \leq C'_{\eps} \times \ptwo{\beta^0 - \beta^0_0} \times \ptwo{\alpha^0 - \alpha^0_0}\]
using Cauchy-Schwarz and the fact that $|\gamma_0 - \gamma| < 1$.

The remaining terms can be bounded in similar fashion to altogether yield the bound:
\begin{align*}
    |\partial^2 f(r)|/C'_{\eps} &\leq \ptwo{\beta - \beta_0} \times \ptwo{\alpha - \alpha_0}\\
    &+\ptwo{\gamma - \gamma_0} \times \ptwo{\alpha - \alpha_0}.
\end{align*}
By Equations \eqref{eq:prog_prop_rate} and \eqref{eq:prog_selec_rate}, this term is bounded above by $\delta_N N^{-1/2}$, and condition 2.3 is therefore satisfied. 

\paragraph{\textbf{Condition 2.4}}
\begin{align*}
    \color{black}{\EOm \Psi(\Om; \theta_0, \eta_0) ^ 2} &= \color{black}{\EOm \big[(\alpha_0(1, 0, \bX) - \alpha_0(1, 1, \bX) - \theta_0) ^ 2 \big]}\\
    &+ \color{black}{\EOm\left(\frac{T(1 - S)(Y - \alpha_0(1, 0, \bX))}{\beta_0(0, \bX)(1 - \gamma_0(\bX))} - \frac{TS(Y - \alpha_0(1, 1, \bX))}{\beta_0(1, \bX)\gamma_0(\bX)}\right) ^ 2} \\
    &\geq \color{black}{\EOm\left(\frac{T(1 - S)(Y - \alpha_0(1, 0, \bX))^2}{\beta^2_0(0, \bX)(1 - \gamma_0(\bX))^2} + \frac{TS(Y - \alpha_0(1, 1, \bX))^2}{\beta^2_0(1, \bX)\gamma^2_0(\bX)}\right)} \\
    &\geq \color{black}{(1 - \eps)^{-4} \EOm[T(1 - S)(Y - \alpha_0(1, 0, \bX))^2 + TS(Y - \alpha_0(1, 1, \bX))^2\big]} \\
    &= \color{black}{(1 - \eps)^{-4} \EOm[T(1 - S)\epsy^2 + TS\epsy^2]} \\
    &=\color{black}{(1 - \eps)^{-4} \EOm[T\epsy^2]} \\
    &= \color{black}{(1 - \eps)^{-4} \EOm[T]\EOm[\epsy^2] = (1 - \eps)^{-4}(c')^2}
\end{align*}
where $c'$ is a positive constant depending on $c$ from Assumption S1 and the marginal expectation of $T$. 

Thus, all of condition 2 is  also satisfied. 

This completes the proof of Theorem \ref{thm:dml_theta}.

\hfill$\square$
\subsubsection{Proof of Theorem \ref{thm:dml_pate}}\label{sec:proof_thm_dml_pate}
\setcounter{theorem}{5}
\begin{theorem}
Given assumptions A1 -- A3 are satisfied, if $\hat{p}(\bX) \rightarrow p(\bX) $ and $\hat{e}(\bX,t,1) \rightarrow e(\bX,t,1) $ then $\hat{\nu}(t)$ is a consistent estimator of $\nu(t)$ i.e. $\hat{\nu}(t) \overset{p}\rightarrow \nu(t).$
Further, if Assumptions A4 and A6 are also satisfied, then $\sqrt{n}(\nuhat(t) - \nu(t)) \overset{d}{\rightarrow} \mathcal{N}(0, \gammasq(t)),$ where $\gammasq(t) = \sigma^2_Y(t) \E\left[\frac{1}{p(\bX)e(\bX,t,1)} \right] + \E\left[ (\nu(\bX,t) - \nu(t))^2\right]$.
In addition, the vanilla estimator for the variance is consistent with the first-stage estimates used in place of the true nuisance parameters,
$\gammasqhat(t) = \frac{1}{n}\sum_{i=1}^N(\Lambdaithat - \nuhat(t))^2 \overset{p}{\rightarrow} \gammasq(t),$
and it follows that an uniformly valid asymptotic $1-\alpha$ confidence interval is $\left[\nuhat(t) \pm \Phi^{-1}(1-\alpha/2)\sqrt{\gammasqhat(t)/n}\right].$
\end{theorem}
\textit{Proof.}
We will focus on the case $t = 1$, for which the estimand $\theta_0$ is $\nu(1) =: \nu_0$, though the case for $t = 0$ is identical. 
Consider the score
\[\Psi = \alpha(\bX) + \frac{ST}{\beta(\bX)\gamma(\bX)}(Y - \alpha(\bX)) - \nu.\]
The true value of the nuisance parameter $\eta$ is $(\alpha_0(\bX), \beta_0(\bX), \gamma_0(\bX))$, where we overload notation so that: 
\begin{itemize}
    \item $\alpha_0(\bX) = \E[Y\mid T=1, S=0, \bX]$
    \item $\beta_0(\bX) = P(T = 1 \mid S = 1, \bX)$
    \item $\gamma_0(\bX)= P(S = 1 \mid \bX)$,
\end{itemize}
and the true value of the parameter of interest $\nu$ is $\nu_0 = \nu(1) = \E[Y(1)]$.

The score is linear in the sense of Equation \eqref{eq:psi_linear}, which allows us to verify Assumptions 3.1 and 3.2 of \cite{chernozhukov2018double} in order to apply Theorem 3.1 and Corollary 3.1.


\subsection*{Verification of Condition 1 for Theorem~\ref{thm:dml_pate}}
\paragraph{\textbf{Condition 1.1}}
Under A1-A3, we show that our score is unbiased:
\begin{align*}
    &\EOm~[\Psi(\Om; \nu_0, \eta_0)] = \\ 
    &\EOm\bigg[\E[Y \mid T = 1, S = 0, \bX] + \frac{ST(Y - \E[Y \mid T = 1, S = 0, \bX])}{\mathrm{P}(T = 1 \mid S = 1, \bX)\mathrm{P}(S = 1 \mid \bX)} - \E[Y(1)] \bigg].
\end{align*}
The first term is $\EOm[\E[Y \mid T = 1, S = 0, \bX]] = \E_{\bX}[\E[Y \mid T = 1, S = 0, \bX]]$. 

For the second term, we apply the expectation $\EOm$ via the iterated expectations $\E_{\bX}\E_{TS|\bX}\E_{Y|ST\bX}$, yielding:
\begin{align*}
    &\EOm \left[\frac{ST(Y - \E[Y \mid T = 1, S = 0, \bX])}{\mathrm{P}(T = 1 \mid S = 1, \bX)\mathrm{P}(S = 1 \mid \bX)}\right] = \\
    &\E_{\bX}\E_{Y|T=1,S=1,\bX}\left[\frac{P(T = 1 \mid S = 1, \bX)P(S = 1 \mid \bX)}{P(T = 1 \mid S = 1, \bX)P(S = 1 \mid \bX)}(Y - \E[Y\mid T = 1, S = 0, \bX])\right] = \\
    &\E_{\bX}\E_{Y|T=1,S=1,\bX}\left[Y - \E[Y\mid T = 1, S = 0, \bX]\right]. 
\end{align*}
Because $\E[Y(1)]$ is a scalar:
\begin{align*}
    \EOm~[\Psi(\Om; \nu_0, \eta_0)] =&~
    \E_{\bX}[\E[Y \mid T = 1, S = 0, \bX]] + \E_{\bX}[\E[Y \mid T = 1, S = 1, \bX]] - \\ 
    &~\E_{\bX}[\E[Y \mid T = 1, S = 0, \bX]] - \E[Y(1)] \\
    =&~\E_{\bX}[\E[Y \mid T = 1, S = 1, \bX]] - \E[Y(1)] \\
    =&~\E_{\bX}[\E[Y(1) \mid T = 1, S = 1, \bX]] - \E[Y(1)] \\
    \stackrel{A2}{=}&~\E_{\bX}[\E[Y(1) \mid S = 1, \bX]] - \E[Y(1)]\\
    \stackrel{A3}{=}&~\E_{\bX}[\E[Y(1) \mid \bX]] - \E[Y(1)]\\
    =&~\E[Y(1)] - \E[Y(1)]\\
    =&~ 0.
\end{align*}
\paragraph{\textbf{Condition 1.2}}

Additionally, assuming A4, we now verify the score is Neyman-orthogonal. For any $\eta \in \mathcal{W}_N$, the G\^{a}teaux derivative in the direction $\eta - \eta_0$ is: 
\begin{align*}
    \partial_\eta \EOm [\Psi(\Om; \nu_0, \eta_0)][\eta - \eta_0] &=\EOm[(\alpha(\bX) - \alpha_0(\bX))] \\ 
    &- \EOm\left[\frac{(\alpha(\bX) - \alpha_0(\bX)))ST}{\beta_0(\bX)\gamma_0(\bX)}\right] \\
    &- \EOm\left[\frac{(Y - \alpha_0(\bX))(\gamma(\bX) - \gamma_0(\bX))ST}{\beta_0(\bX)\gamma_0^2(\bX)}\right] \\
    &- \EOm\left[\frac{(Y - \alpha_0(\bX))(\beta(\bX) - \beta_0(\bX))ST}{\gamma_0(\bX)\beta_0^2(\bX)}\right]. 
\end{align*}
The first two terms cancel, as:
\begin{align*}
    &\EOm[(\alpha(\bX) - \alpha_0(\bX))] - \EOm\left[\frac{(\alpha(\bX) - \alpha_0(\bX))ST}{\beta_0(\bX)\gamma_0(\bX)}\right] = \\ &\E_{\bX}\E_{TS|X}\left[(\alpha(\bX) - \alpha_0(\bX))\left(1 - \frac{ST}{\beta_0(\bX)\gamma_0(\bX)}\right)\right] = \\
    &\E_{\bX}\left[(\alpha(\bX) - \alpha_0(\bX))\E_{TS|X}\left[1 - \frac{ST}{\beta_0(\bX)\gamma_0(\bX)}\right]\right] = \\
    &\E_{\bX}\left[(\alpha(\bX) - \alpha_0(\bX))\E_{TS|X}\left[1 - \frac{\beta_0(\bX)\gamma_0(\bX)}{\beta_0(\bX)\gamma_0(\bX)}\right]\right] = 0.
\end{align*}
Similarly, rewriting the third term:
\begin{align*}
&\EOm\left[\frac{(Y - \alphuxz)(\gamux - \gamuxz)ST}{\betuxz\gamma_0^2(\bX)}\right] = \\
&\E_{\bX}\left[\left(\frac{\gamux - \gamuxz}{\gamux}\right)\E_{TS|X}\E_{Y|TSX}\left[\frac{(Y - \alphuxz)ST}{\betuxz\gamma_0(\bX)}\right]\right] = \\
&\E_{\bX}\left[\left(\frac{\gamux - \gamuxz}{\gamux}\right)\E_{Y|T=1,S=1,\bX}(Y - \alphuxz)\right] = \\
&\E_{\bX}\bigg[\left(\frac{\gamux - \gamuxz}{\gamux}\right)(\E[Y \mid T = 1, S = 1, \bX] - \E[Y \mid T = 1, S = 0, \bX])\bigg] = \\
&\E_{\bX}\bigg[\left(\frac{\gamux - \gamuxz}{\gamux}\right)(\E[Y(1) \mid T = 1, S = 1, \bX] - \E[Y(1) \mid T = 1, S = 0, \bX])\bigg] \stackrel{A2,A4}{=} \\
&\E_{\bX}\bigg[\left(\frac{\gamux - \gamuxz}{\gamux}\right)(\E[Y(1) \mid S = 1, \bX] - \E[Y(1) \mid S = 0, \bX])\bigg] \stackrel{A3}{=} \\
&\E_{\bX}\bigg[\left(\frac{\gamux - \gamuxz}{\gamux}\right)(\E[Y(1) \mid \bX] - \E[Y(1) \mid \bX])\bigg] = \\
&0.
\end{align*}

Analogous reasoning about the fourth term, shows that it, and therefore the G\^{a}teaux derivative map are 0. 
The fact that the score is unbiased under A3 alone, but that A4 is required for Neyman-Orthogonality, reflects the fact that efficient estimation is possible only under (A3, A4), whereas consistency is achieved even under A3.
\paragraph{\textbf{Conditions 1.2 -- 1.5}}
We have already shown the score to be linear and the map $\eta \rightarrow \EOm [\Psi]$ can trivially be shown to be twice G\^{a}teaux differentiable, satisfying conditions 1.2 to 1.4. Lastly, because $\Psi^a = -1$, the identification requirement of condition 1.5 -- that the singular values of $\EOm[\Psi^a]$ (here, 1) are bounded between $c_0$ and $c_1$ -- is satisfied as well. 

Thus, all of condition 1 is satisfied for proof of theorem~\ref{thm:dml_pate}. 

\subsection*{Verification of Condition 2 for Theorem \ref{thm:dml_pate}}

\paragraph{\textbf{Condition 2.1}}
This is automatically satisfied by Assumption S1(g) and the construction of $\mathcal{W}_N$.

\paragraph{\textbf{Condition 2.2}}
\color{black}
We now need to bound the $L_q$ norms of $\Psi$ and $\Psi^a$. To do so, first rewrite the score:

\[
\Psi(\Om; \nu, \eta) = \alphux\left(1 - \frac{ST}{\betux\gamux}\right) + \frac{STY}{\betux\gamux} - \nu
\]

We can then bound $\pq{\Psi(\Om; \nu, \eta_0)}$:
\begin{equation}\label{eq:psi_nu_norm}
\begin{split}
    \pq{\Psi(\Om; \nu_0, \eta)} \leq & \pq{\alphux\left(1 - \frac{ST}{\betux\gamux}\right)} + \\
    & \pq{\frac{STY}{\betux\gamux}} + |\nu_0| \\
    \leq & (1 + \eps^{-2}) \pq{\alphux} + \eps^{-2}\pq{Y} + |\nu_0| \\
    \leq & (1 + \eps^{-2}) \pq{\alphux - \alphuxz + \alphuxz} + \eps^{-2}\pq{Y} + |\nu_0| \\
    \leq & (1 + \eps^{-2}) \pq{\alphux - \alphuxz} + (1 + \eps^{-2})\pq{\alphuxz} + \eps^{-2}\pq{Y} + |\nu_0| \\
\end{split}    
\end{equation}

where we use the fact that $S$ and $T$ are binary and that $\betux$ and $\gamux$ are bounded below by $\eps$. We can now bound each of the terms of \eqref{eq:psi_nu_norm} in turn. 

The first two terms are bounded by Lemma \ref{lemma:norm_bounds} and the third by Assumption S1(b), so it remains to bound $|\nu_0|$:
\[|\nu_0| = |\E[Y(1)]|= |\E[\E[Y(1) \mid \bX]]| = |\E \alpha_0(\bX)| \leq \ptwo{\alpha_0(\bX)} \leq C/\eps\]
where the inequalities follow by Jensen's inequality and Lemma \ref{lemma:norm_bounds}. 

Thus, for any $\eta \in \mathcal{W}_N$, we have that
\begin{align*}
    \pq{\Psi(\Om; \nu_0, \eta)} \leq 2C(1 + \eps^{-2})/\eps^{2/q} + C/\eps ^ 2 + C/\eps
\end{align*}
giving the required bound on $m_N$ in Assumption 2.2.. Lastly, because $\pq{\Psi^a} = 1$, the required bound on $m_N'$ is also satisfied. Thus, Condition 2.2 is satisfied. 
\paragraph{\textbf{Condition 2.3}}
Because $\Psi^a$ does not involve nuisance parameters,
\[\|\EOm[\Psi^a(\Om; \eta) - \Psi^a(\Om; \eta_0)]\| = |-1 + 1| = 0 \leq \delta_N\]
and so the bound on the rate $r_N$ is satisfied. 

Next, we bound the rate $r'_N$. By the triangle inequality, 
\[\ptwo{\Psi(\Om; \nu_0, \eta) - \Psi(\Om; \nu_0, \eta_0)} \leq \mathcal{I}_1 + \mathcal{I}_2\]
for 
\begin{gather*}
    \mathcal{I}_1 := \ptwo{\alphux - \alphuxz} \\
    \mathcal{I}_2 := \left\Vert \frac{TS(Y - \alphux)}{\gamux\betux} - \frac{TS(Y - \alphuxz)}{\gamuxz\betuxz} \right\Vert_{P, 2}
\end{gather*}
By Lemma \ref{lemma:norm_bounds}, $\mathcal{I}_1$ is bounded by $\delta_N/\eps$.

To bound $\mathcal{I}_2$, we combine the fractions and bound the denominator by $\eps^4$ to yield
\begin{align*}
    \mathcal{I}_2 &\leq \eps^{-4} \ptwo{TS(Y - \alphux)\gamuxz\betuxz - TS(Y - \alphuxz)\gamux\betux} \\
    &\leq \eps^{-4} \ptwo{(\alphuxz + \epsy - \alphux)\gamuxz\betuxz - \epsy\gamux\betux} \\
    &\leq \eps^{-4} \left(\ptwo{(\alphuxz - \alphux)\gamuxz\betuxz} + \ptwo{\epsy(\gamuxz\betuxz - \gamux\betux)}\right) \\
    &\leq \eps^{-4} \left(\ptwo{(\alphuxz - \alphux)} + \sqrt{C}\ptwo{(\gamuxz\betuxz - \gamux\betux)}\right) \\
    &\leq \eps^{-4}(\eps^{-1} + \sqrt{C}C_{\delta})\delta_N
\end{align*}
where $C_{\delta}$ is a constant only depending on the first term of the sequence $(\delta_n)_{n=1}^\infty$. In the second inequality we use Equation \eqref{eq:alpha0_eq}. In the third inequality, we apply the triangle inequality, in the fourth inequality we use the fact that $\gamuxz\betuxz < 1$ and Assumption S1, and in the fifth inequality we use Lemma \ref{lemma:norm_bounds} and the facts that $\gamuxz\betuxz < 1$ and $\gamux\betux < 1$.
Using the bounds for $\mathcal{I}_1, \mathcal{I}_2$, we can then bound
\[\ptwo{\Psi(\Om; \nu_0, \eta) - \Psi(\Om; \nu_0, \eta_0)} \leq \delta_N(\eps^{-1} + \eps^{-5} + \sqrt{C}C_\delta \eps^{-4}) \leq \delta_N'\] as long as $K$ in the definition of $\delta'_N$ satisfies $K \geq \eps^{-1} + \eps^{-5} + \sqrt{C}C_\delta \eps^{-4}$. Thus, we have the requisite bound on $r'_N$.

Next, we need to bound $|\partial^2 f(r)|$, where $f(r) := \EOm[\Psi(\Om; \nu_0, \eta_0 + r(\eta - \eta_0))]$ for $r \in (0, 1)$. For any $r \in (0, 1)$:
\begin{align*}
\partial^2 f(r) &= \\
    &\EOm\bigg[\frac{2 S T (\betux-\betuxz)^2 
  (-r (\alphux-\alphuxz)-\alphuxz+Y)}{(r (\betux-\betuxz)+\betuxz)^3 
    (r (\gamux-\gamuxz )+\gamuxz )}\bigg] + \\
&\EOm\bigg[\frac{2 S T (\betux-\betuxz) (\gamux-\gamuxz ) 
      (-r (\alphux-\alphuxz)-\alphuxz+Y)}{(r (\betux-\betuxz)+\betuxz)^2 
        (r (\gamux-\gamuxz )+\gamuxz )^2}\bigg] + \\
&\EOm\bigg[\frac{2 S T (\gamux-\gamuxz )^2 
          (-r (\alphux-\alphuxz)-\alphuxz+Y)}{(r (\betux-\betuxz)+\betuxz) 
            (r (\gamux-\gamuxz )+\gamuxz )^3}\bigg] - \\
&\EOm\bigg[\frac{2 S T (\alphuxz-\alphux) 
              (\betux-\betuxz)}{(r (\betux-\betuxz)+\betuxz)^2 (r (\gamux-\gamuxz )+\gamuxz )}\bigg] - \\
&\EOm\bigg[\frac{2 S T (\alphuxz-\alphux) (\gamux-\gamuxz )}{(r (\betux-\betuxz)+\betuxz)
  (r (\gamux-\gamuxz )+\gamuxz )^2}\bigg].
\end{align*}

To begin to bound $|\partial^2 f(r)|$, note that the expectation and the absolute value can be exchanged by Jensen's inequality, and the absolute value then distributed across the sum. Then, note that the absolute value of each denominator has an infimum over $(r, \bX)$ that is bounded away from 0\footnote{Each denominator is a polynomial whose absolute value is bounded away from 0 because each contains at least one term that is at most a function of $\betux$, $\betuxz$, $\gamux$, and $\gamuxz$, each of which are bounded away from 0.} and so we can bound one over the absolute value of the denominator of each term all by one over the minimum of the infima of the absolute values of the denominators, which will be a constant depending solely on $\epsilon$. In the below, we thus use $C'_{\epsilon}$ to denote a constant (possibly changing from line to line) that depends at most on $C$ and $\epsilon$. 

The absolute value of the fourth term can be bounded by:
\begin{align*}
    &C'_{\eps}\EOm\big[2 S T \left\lvert(\alphuxz-\alphux) 
              (\betux-\betuxz)\right\rvert\big] \leq \\
    &C'_{\eps}\EOm\big[\left\lvert(\alphuxz-\alphux) 
              (\betux-\betuxz)\right\rvert\big] \leq \\
    &C'_{\eps}\|\alphuxz-\alphux\|_{P, 2} \times \|\betux-\betuxz\|_{P, 2}
\end{align*}
\color{black}
where we use the fact that $S$ and $T$ are binary and Cauchy-Schwarz. The fifth term can in an analogous fashion be bounded by:
\[C'_{\eps}\|\alphuxz-\alphux\|_{P, 2} \times \|\gamux-\gamuxz\|_{P, 2}.\]
The absolute value of the first term can be bounded by: 
\begin{align*}
  &C'_{\eps}\EOm 2ST(\betux-\betuxz)^2 \left\lvert
  (-r (\alphux-\alphuxz)-\alphuxz+Y)\right\rvert = \\
  &C'_{\eps}\EOm\bigg[2 S T |(\betux-\betuxz)^2 
  (-r (\alphux-\alphuxz) + \epsy)|\bigg] = \\
  &C'_{\eps}\left(\EOm\big[S T (\betux-\betuxz)^2 |\epsy|\big] + 
  \EOm\big[ST|(\betux - \betuxz) ^ 2 (r (\alphux-\alphuxz))|\big] \right) \leq \\
  &C'_{\eps}\left(\EOm\big[S T  |\epsy|\big] + 
  \EOm\big[ST|(\betux - \betuxz) (r (\alphux-\alphuxz))|\big] \right) \leq \\
  &C'_{\eps}\EOm\big[|(\betux - \betuxz) (\alphux-\alphuxz))|\big] \leq \\
  &C'_{\eps}\|\alphuxz-\alphux\|_{P, 2} \times \|\betux-\betuxz\|_{P, 2}.
\end{align*}
In the first equation, we use Equation \eqref{eq:alpha0_eq}. In the first inequality we use the fact that $|\betux - \betuxz| \leq 2$. In the second inequality we use the facts that $S$ and $T$ are binary, that $r$ is a constant less than 1, and Assumption S1. The final inequality follows by Cauchy-Schwarz. 
The third term can analogously be bounded by:

\[C'_{\eps}\|\alphuxz-\alphux\|_{P, 2} \times \|\gamux-\gamuxz\|_{P, 2}\]      
and the second by:
\begin{align*}
    &C'_{\eps}\EOm\big[|(\gamux - \gamuxz)(\betux - \betuxz) (\alphux-\alphuxz))|\big] \\
    \leq &C'_{\eps}\EOm\big[|(\betux - \betuxz) (\alphux-\alphuxz))|\big] \\
    \leq &C'_{\eps}\|\alphuxz-\alphux\|_{P, 2} \times \|\betux-\betuxz\|_{P, 2} 
\end{align*}
where in the first inequality we use the fact that $|\betux - \betuxz| \leq 2$.

\noindent Putting the pieces together, it follows that:
\begin{align*}
|\partial^2 f(r)|/C'_\eps 
&\leq \ptwo{\gamux - \gamuxz} \times \ptwo{\alphux - \alphuxz} \\
& +\ptwo{\betux - \betuxz} \times \ptwo{\alphux - \alphuxz} 
\end{align*}
and so condition 2.3 is satisfied. 

\paragraph{\textbf{Condition 2.4}}
We require the eigenvalues of $$\EOm[\Psi(\Om; \nu_0, \eta_0)\Psi(\Om; \nu_0, \eta_0)']$$ to be bounded from below by some $c_0 > 0$.

\begin{align*}
    \EOm \Psi ^ 2(\Om;\theta_0,\eta_0) &=\EOm (\alphuxz - \nu_0) ^ 2 + \EOm\left(\frac{ST}{\betuxz\gamuxz}(Y - \alphuxz)\right) ^ 2 \\
    &\geq \EOm\left(\frac{ST}{\betuxz\gamuxz}(Y - \alphuxz)\right) ^ 2 \\
    &= \EOm\left(\frac{ST}{\beta^2_0(\bX) \gamma^2_0(\bX)}(Y - \alphuxz) ^ 2\right) \\
    &\geq \frac{1}{(1 - \eps) ^ 4} \EOm(ST(Y - \alphuxz)) ^ 2\\
    &= \frac{1}{(1 - \eps) ^ 4} \EOm(\epsy) ^ 2 \\
    &\geq \frac{c ^ 2}{(1 - \eps) ^ 4}
\end{align*}
where the cross-term in the first equality goes to zero under Equation \eqref{eq:alpha0_eq} by the same calculations used to verify \textit{Score Unbiasedness} and the final inequality follows from Assumption S1. 

This verifies both conditions and therefore completes the proof of Theorem \ref{thm:dml_pate}.
\hfill$\square$

\section{Coronary Artery Surgery Study (CASS)}\label{sec:cass}
\subsection{Data Description}
The coronary artery surgery study (CASS) was initiated by National Heart, Lung and Blood institute (NHLBI) to study the effect of coronary bypass surgery in comparison to conventional medical therapy. The data were collected from 15 medical centers across the US and Canada from 1974 to 1979, yielding 24,989 patients.
Of these, 2,099 eligible patients were selected for the randomized control trial (RCT) and were part of a comprehensive follow-up study. 780 of the 2,099 patients accepted the randomized treatment -- we refer to them as the experimental arm in our analysis. 1,319 patients refused the randomization and self-selected into treatment groups -- this is our observational arm. For our analysis, in the experimental arm, the treatment group is defined based on intent-to-treat by the original randomized assignment. Note that $23.5\%$ of patients assigned to medical therapy had bypass surgery within 5 years since their angina worsened. In the observational arm, a similar treatment group was identified as any patient who was selected for surgery within 90 days of enrollment or if their surgery was in the first year period (when 95\% of CASS experimental arm surgeries were done). Any observational study patient who did not have early elective surgery were treated as the control group (or medical therapy arm). Here, we use all-cause mortality during the course of the study as the outcome of interest.

\subsection{Analysis and Result}
In this paper, we are interested in studying if the conditional ignorability of the observational sample as well as the external validity of the experimental sample holds. Furthermore, we are also interested in estimating the population average treatment effect of surgical intervention as compared to medical therapy. 

Given that the same set of practitioners administers the treatments for both the observational as well as experimental samples, it is reasonable to assume that selection in the experimental (or observational) arm does not have a direct causal link to the outcome. That implies that Assumption A5 is likely to hold. 

Our test allows us to study if either A3 or/and A4 are violated. If we find significant evidence for $\theta(t)\neq0$ for any $t\in\{0,1\}$, then either of these assumptions can be violated. The result is that our test \textit{fails to reject the null hypothesis} and finds \textit{no  evidence} for the violation of these assumptions. The p-values corresponding to the tests are 0.290 and 0.915, respectively for $\theta(0)$ and $\theta(1)$. We also show the distributions of $\theta(0)$ and $\theta(1)$ in Figure~\ref{fig:theta_cass}.
\begin{figure}
    \centering
    \includegraphics[width=0.6\textwidth]{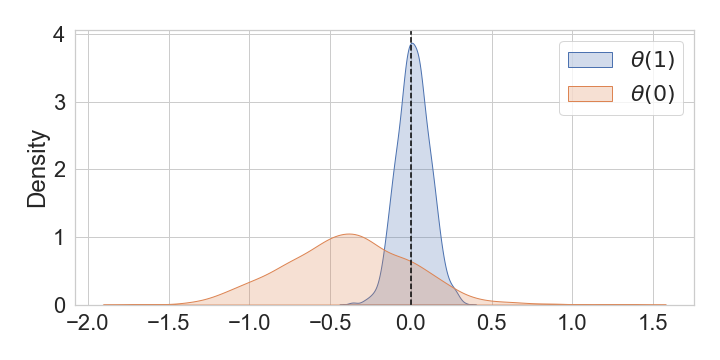}
    \caption{Kernel density plot for $\theta(0)$ and $\theta(1)$ for the CASS dataset. }
    \label{fig:theta_cass}
\end{figure}
Note that failure to find evidence does not necessarily imply that there is no unobserved confounding. However, these results are in congruence with previous works \citep{dahabreh2020extending,olschewski1992analysis}, suggesting that the experimental sample had external validity and the observational sample has conditional ignorability.

Next, we estimate the average treatment effect using our estimator and other estimators. Table~\ref{tab:cass_ate} presents the estimated average treatment effects using our estimator proposed in Equation~\eqref{eq:nuhatdml},  the difference in means estimator for the experimental sample and an augmented inverse propensity score weighted (AIPW) estimator in the observational sample. Since our tests failed to reject, it is not surprising that the three estimates are effectively the same and consistent with the literature: the treatment effect is not significantly different than 0, i.e., the coronary bypass surgery neither helps nor hurts patients' chances of survival. 

\begin{table}
    \caption{\label{tab:cass_ate} ATE Estimates CASS data using different approaches on experimental and observational samples.}
    \centering
    \begin{tabular}{lrr}
    \toprule
    {} &    \textbf{Estimated ATE} &    \textbf{Std. Error} \\
    \midrule
    Our ATE Est.       & -0.019 &  0.024 \\
    Exp. ATE Est.      & -0.013 &  0.032 \\
    Obs. ATE Est.      & -0.022 &  0.034 \\
    \bottomrule
    \end{tabular}
\end{table}
\section{Lalonde Data: Evaluating Training Programs}\label{sec:lalonde}
\subsection{Data Description}
The Lalonde data consist of a randomized control trial, the National Support Work Demonstration (NSW), that studied the effect of training programs on participant income levels \citep{lalonde1986}. The NSW is often augmented with the Population Survey of Income Dynamics (PSID-2) dataset of observational control units \citep{dehejia1999}; together, the two can serve as a benchmark for observational causal inference methods. While most estimation methods struggle with recovering the experimental point estimate using the joint dataset, some methods are able to recover the experimental ATE after pre-processing \citep[e.g.,][]{malts}. We reanalyze the NSW and PSID-2 datasets to study if the experimental controls are comparable to observational controls; that is, if the experiment satisfies external validity (Assumption A3).
We limit our analysis to the subsample of male household heads under the age of 55 and who are not retired by 1975. The outcome of interest is the income of participants in 1978. Furthermore, for both datasets, we use pre-treatment information about units' age, race, marital status, education, and income in 1975. 

\subsection{Analysis and Results}
Unlike our analyses of the STAR and CASS datasets, the observational data in this analysis consists only of control units. Our analysis is thus limited to identifying a possible violation of experimental external validity (A3) by testing if $\theta(0)=0$. Our test rejects the null hypothesis (p-value = $6.2 \times 10^{-6}$), indicating that the NSW experiment lacks external validity (with respect to PSID-2). As such, it is not at all surprising that many observational causal inference approaches are unable to recover experimental ATE when using the observational control group. One notable exception to this behavior is using the matching algorithm MALTS \citep{malts}.

We reanalyze the joint NSW and PSID-2 data using MALTS, a matching algorithm that learns a distance metric to guarantee tighter matches on more important covariates. In this approach, we estimate a matched group of size 10 for each unit in the sample and calculate a diameter of the matched group with respect to the learned metric. We plot these diameters in Figure~\ref{fig:prune_lalonde} and note the large gap between diameters of size 80 and 100. Using this heuristic we prune (eliminate) units from the analysis that have a diameter that's larger than 80; we emphasize that we do \textit{not} use outcome information in order to prune.
It is important to note that the pruned set has a significant number of control units from the observational arm along with units from the experimental arm as shown in Figure~\ref{fig:prune_lalonde} by the color of the marker. 

After pruning, the test for violation of A3 then \textit{fails} to reject the null (p-value = $0.79$ and Figure~\ref{fig:theta_lalonde} shows the point estimate of $\theta_0$ with respect to the reference null distribution before and after pruning). While the matched groups were pruned only based on the tightness of the match, the corresponding change in the test statistic suggests that it is possible that an  unobserved confounder causing selection in the experiment is actually \textit{correlated} with observed pre-treatment covariates. This provides a compelling explanation for why most observational approaches fail to recover the experimental ATE using an observational arm, while a flexible matching framework is able to do so. 

\begin{figure}
    \centering
    \includegraphics[width=0.6\textwidth]{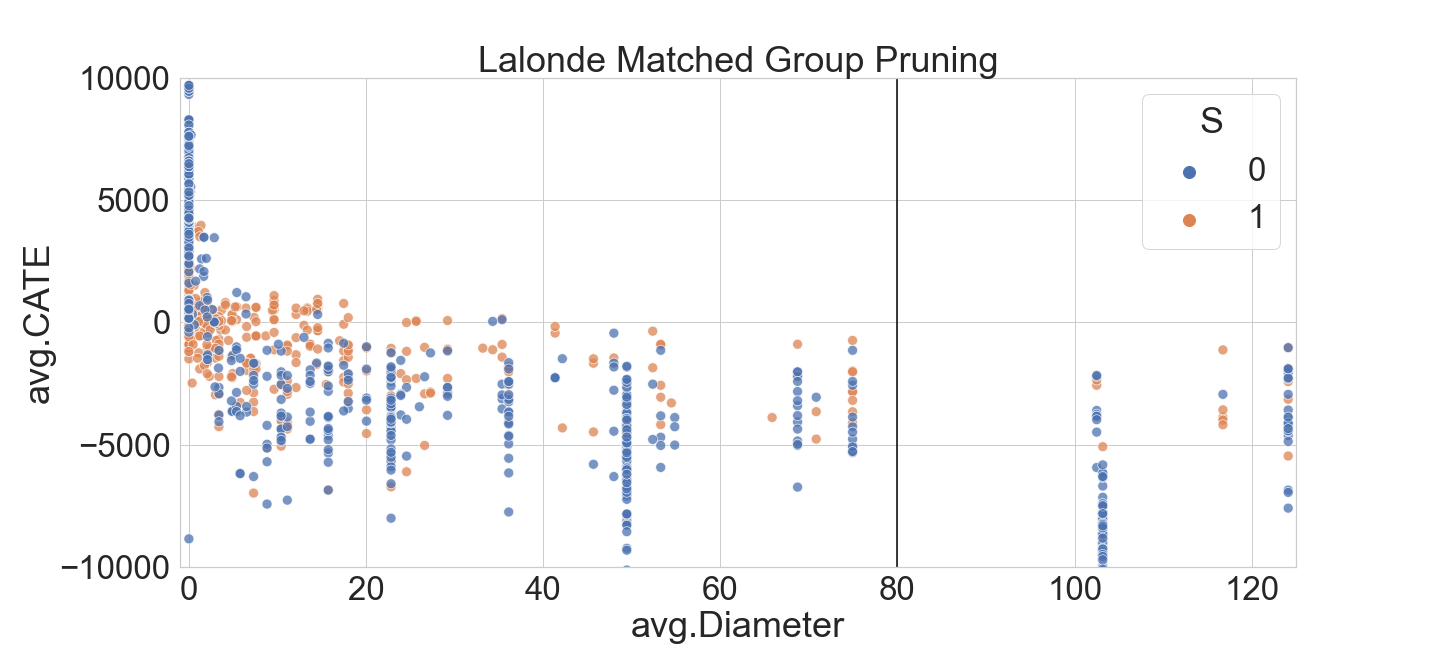}
    \caption{Pruning criteria for matched groups created by Lalonde based on the tightness of matches (measured as the diameter of the matched group). The black vertical line at 80 denotes the threshold above which matched groups are eliminated because the control and treated units are no longer well-matched.}
    \label{fig:prune_lalonde}
\end{figure}

\begin{figure}
    \centering
    \includegraphics[width=0.6\textwidth]{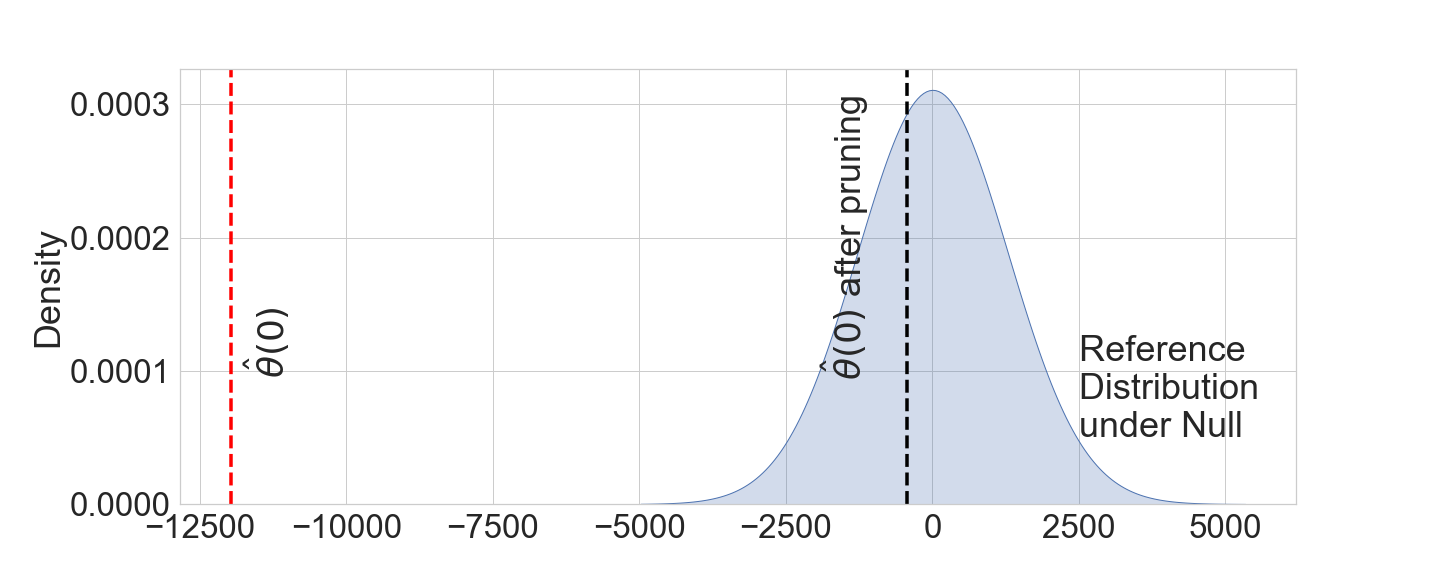}
    \caption{Distribution of the null distribution (blue) and point estimate of $\hat{\theta}(0)$ (red) in the Lalonde data.}
    \label{fig:theta_lalonde}
\end{figure}
\section{Synthetic Experiments}\label{sec:syn_exp}
\subsection{Efficiency Study}
\begin{figure}
    \centering
    \includegraphics[width=0.9\textwidth]{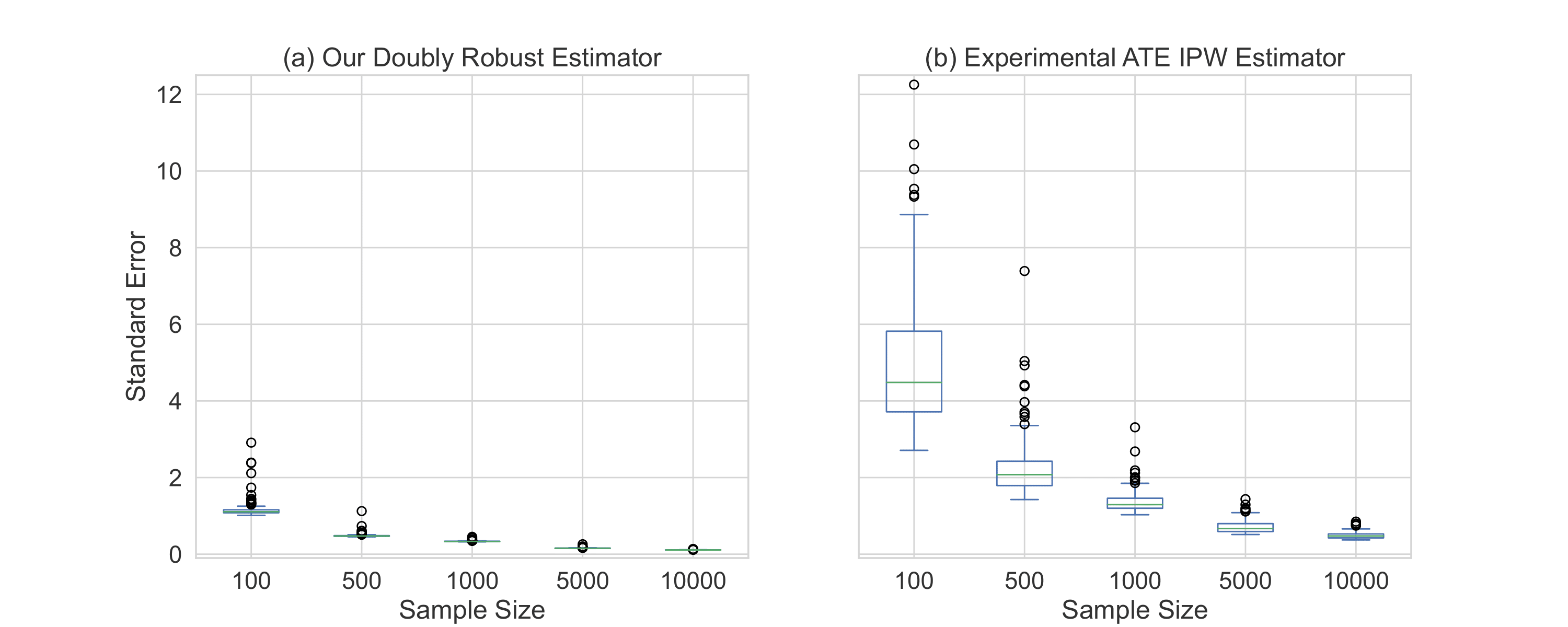}
    \caption{Boxplots of standard error for ATE estimates using (a) our doubly robust estimator and (b) IPW estimator on experimental data generated over 500 simulations for each sample-size.}
    \label{fig:efficiency}
\end{figure}

In this subsection, we study and compare the variance estimates for different ATE estimators. Particularly, we are interested in scenario when the experimental sample has external validity and the experimental SATE is generalizable. Here, we compare the ATE estimates and associated variance for (a) our proposed doubly robust estimator, and (b) IPW experimental ATE estimate. For our analysis, we use synthetic data experiments generated as follows:
\begin{eqnarray*}
  \{X_{i,1}, X_{i,2}, X_{i,3}, X_{i,4}\} &\overset{iid}{\sim}& \mathcal{N}(0,1) \\
  \pi_{S,i} &=& X_{i,1} - X_{i,2} \\
  \pi^{exp}_{T,i} &=& 0.5 \\
  \pi^{obs}_{T,i} &=& X_{i,1} + X_{i,2} - 2 * X_{i,3}  \\
  S_i &\sim& Bernoulli(\pi_{S,i}) \\
  T_i &\sim& Bernoulli( S_i \pi^{exp}_{T,i} + (1-S_i) \pi^{obs}_{T,i})\\
  Y_i &=& 5 X_{i,2} + X_{i,4} + T_i (X_{i,1} + X_{i,3}).
\end{eqnarray*}
For each sample size, we first generate covariates for each unit $i$: $X_{i,1} \dots X_{i,4}$. We then simulate 500 different versions of sample and treatment assignments as a function of the covariates. For each sample, we estimate then estimate ATEs and respective standard errors using our estimators of interest. Figure~\ref{fig:efficiency} shows the distribution of standard errors for each sample size. We observe that for both the estimators the standard errors shrink as the sample size increases. Further, the standard errors for our doubly robust estimator are consistently smaller than IPW experimental sample ATE estimator.

\end{document}